%% file: B2G-20-009_temp.tex
\pdfoutput=1
\documentclass[11pt,twoside,a4paper,cmspaper,final,collab]{cms-tdr}
\def\svnVersion{8185cb0}\def\svnDate{2023/07/24}\def\cmsCernNoTag{CERN-EP-2022-152}\def\cmsCernDate{\today}\def\cmsMessage{Published in Physics Letters B as \href{http://dx.doi.org/10.1016/j.physletb.2023.137813}{\doi{10.1016/j.physletb.2023.137813}.}}
\begin{document}\cmsNoteHeader{B2G-20-009}

\newcommand{\INTLUMIsix}{36.3\xspace}
\newcommand{\INTLUMIseven}{41.5\xspace}
\newcommand{\INTLUMIeight}{59.7\xspace}
\newcommand{\INTLUMI}{\ensuremath{138\fbinv}\xspace}
\newcommand{\BulkGMassMin}{\ensuremath{1.3\TeV}\xspace}
\newcommand{\BulkGMassMinNoTeV}{\ensuremath{1.3}\xspace}
\newcommand{\BulkGMassMax}{\ensuremath{6\TeV}\xspace}
\newcommand{\MassExclVPr}{\ensuremath{4.8\TeV}\xspace}
\newcommand{\XsecExclVBF}{\ensuremath{0.1\unit{fb}}\xspace}
\newcommand{\MassExclBulkG}{1.4\xspace}
\newcommand{\MassExclBulkR}{\ensuremath{2.7\TeV}\xspace}
\newcommand{\BulkG}{\ensuremath{\PXXG_{\text{bulk}}}\xspace}
\newcommand{\Radion}{\ensuremath{\text{Rad}\xspace}}
\newcommand{\Mthr}{1250\GeV\xspace}
\newcommand{\MVV}{\ensuremath{m_\mathrm{jj}^{\mathrm{AK8}}}\xspace}
\newcommand{\MJ}{\ensuremath{m_\text{jet}^{\mathrm{AK8}}}\xspace}
\newcommand{\MJO}{\ensuremath{m_\text{jet1}^{\mathrm{AK8}}}\xspace}
\newcommand{\MJT}{\ensuremath{m_\text{jet2}^{\mathrm{AK8}}}\xspace}
\newcommand{\MJOT}{\ensuremath{m_\text{jet1/2}^{\mathrm{AK8}}}\xspace}
\newcommand{\MJTO}{\ensuremath{m_\text{jet2/1}^{\mathrm{AK8}}}\xspace}
\newcommand{\DeepWZqq}{\ensuremath{\qqbar}\xspace}
\newcommand{\PVpr}{\ensuremath{\PV^\prime}\xspace}
\newcommand{\kappatilde}{\ensuremath{\widetilde{\kappa}}\xspace}
\newcommand{\rc}{\ensuremath{r_\mathrm{c}}\xspace}
\newcommand{\lambdaR}{\ensuremath{\Lambda_\mathrm{R}}}\xspace
\newcommand{\cF}{\ensuremath{c_\mathrm{F}}\xspace}
\newcommand{\cH}{\ensuremath{c_{\PH}}\xspace}
\newcommand{\gV}{\ensuremath{g_\PV}\xspace}
\newcommand{\DeepZHbb}{\ensuremath{\bbbar}\xspace}
\newcommand{\VV}{\ensuremath{\PV\PV}\xspace}
\newcommand{\VH}{\ensuremath{\PV\PH}\xspace}
\newcommand{\Vjets}{\ensuremath{\PV\text{+jets}}\xspace}
\newcommand{\Zjets}{\ensuremath{\PZ\text{+jets}}\xspace}
\newcommand{\Wjets}{\ensuremath{\PW\text{+jets}}\xspace}
\newcommand{\WZ}{\ensuremath{\PW\PZ}\xspace}
\newcommand{\ZH}{\ensuremath{\PZ\PH}\xspace}
\newcommand{\WH}{\ensuremath{\PW\PH}\xspace}
\newcommand{\ZZ}{\ensuremath{\PZ\PZ}\xspace}
\newcommand{\WW}{\ensuremath{\PW\PW}\xspace}
\newcommand{\pp}{\ensuremath{\Pp\Pp}\xspace}
\newlength\cmsTabSkip\setlength{\cmsTabSkip}{1ex}

\cmsNoteHeader{B2G-20-009}

\title{Search for new heavy resonances decaying to \texorpdfstring{\WW}{WW}, \texorpdfstring{\WZ}{WZ}, \texorpdfstring{\ZZ}{ZZ}, \texorpdfstring{\WH}{WH}, or \texorpdfstring{\ZH}{ZH} boson pairs in the all-jets final state in proton-proton collisions at \texorpdfstring{$\sqrt{s}=13\TeV$}{sqrt(s) = 13 TeV}}

\date{\today}

\abstract{
A search for new heavy resonances decaying to \WW, \WZ, \ZZ, \WH, or \ZH boson pairs in the all-jets final state is presented. The analysis is based on proton-proton collision data recorded by the CMS detector in 2016--2018 at a centre-of-mass energy of 13\TeV at the CERN LHC, corresponding to an integrated luminosity of \INTLUMI. The search is sensitive to resonances with masses between \BulkGMassMinNoTeV and \BulkGMassMax, decaying to bosons that are highly Lorentz-boosted such that each of the bosons forms a single large-radius jet. Machine learning techniques are employed to identify such jets. No significant excess over the estimated standard model background is observed. A maximum local significance of 3.6 standard deviations, corresponding to a global significance of 2.3 standard deviations, is observed at masses of 2.1 and 2.9\TeV. In a heavy vector triplet model, spin-1 \PZpr and \PWpr resonances with masses below \MassExclVPr are excluded at the 95\% confidence level (\CL). These limits are the most stringent to date. In a bulk graviton model, spin-2 gravitons and spin-0 radions with masses below \MassExclBulkG and \MassExclBulkR, respectively, are excluded at 95\% \CL. Production of heavy resonances through vector boson fusion is constrained with upper cross section limits at 95\% \CL as low as \XsecExclVBF.
}

\hypersetup{
pdfauthor={CMS Collaboration},
pdftitle={Search for new heavy resonances decaying to WW, WZ, ZZ, WH, or ZH boson pairs in the all-jets final state in proton-proton collisions at sqrt(s) = 13 TeV},
pdfsubject={CMS},
pdfkeywords={CMS, diboson resonances, substructure}}

\maketitle

\section{Introduction}
\label{sec:Introduction}

The CERN LHC allows the probing of new phenomena in interactions of elementary particles at energies of multiple \TeVns.
While the standard model (SM) of particle physics describes these high-energy interactions very successfully, it leaves several questions unresolved, such as the nature of dark matter and the origin of the large difference between the electroweak and Planck scales.
Theories beyond the SM that can address these questions introduce new particles and interactions that could be observed in proton-proton (\pp) collisions at the LHC.
A wide range of models predict the production of new heavy resonances decaying to pairs of \PW, \PZ (jointly referred to as \PV), and Higgs bosons (\PH). Examples of such resonances are spin-0 radions (\Radion) and spin-2 gravitons (\BulkG) in the Randall--Sundrum model with warped extra dimensions~\cite{Randall:1999ee,Randall:1999vf,Goldberger:1999un,Goldberger:1999uk,Agashe:2007zd, Fitzpatrick:2007qr, Antipin:2007pi}, and spin-1 vector boson resonances (\PWpr and \PZpr)~\cite{Grojean:2011vu,Salvioni:2009mt} appearing in composite Higgs~\cite{Bellazzini:2014yua,CHM2,Composite2,Greco:2014aza,Lane:2016kvg} and little Higgs~\cite{Schmaltz:2005ky,ArkaniHamed:2002qy} models and forming a heavy vector triplet (HVT)~\cite{Pappadopulo:2014qza}.
Previous searches by the CMS Collaboration in the \VV~\cite{CMS:2021klu, Sirunyan:2019jbg,Sirunyan:2019vgt,Sirunyan:2018iff,Sirunyan:2018hsl,Sirunyan:2018ivv,Sirunyan:2017jtu,Sirunyan:2017acf,Sirunyan:2016cao,Khachatryan:2014xja,Khachatryan:2014hpa,Khachatryan:2014gha} and \VH~\cite{CMS:2021klu, Sirunyan:2018fuh,Sirunyan:2018qob,Sirunyan:2017wto,Khachatryan:2016cfx,Khachatryan:2016yji,Khachatryan:2015bma,Khachatryan:2015ywa} channels, and corresponding searches by the ATLAS Collaboration in the \VV~\cite{Aad:2020ddw,Aad:2019fbh,Aaboud:2018bun,Aaboud:2017ahz,Aaboud:2017eta,Aaboud:2016okv,Aad:2015ufa,Aad:2014xka,ATLASwprimeWZPAS} and \VH~\cite{Aad:2020tps,Aad:2015yza,ATLASVH13,Aad:2015owa} channels, have not observed significant deviations from the SM.
The most stringent lower limits at 95\% confidence level (\CL)~\cite{CMS:2021klu,Aad:2020ddw} for spin-1 resonances decaying to a \WZ (\WH) boson pair exclude masses up to 4.3 (4.0)\TeV in the HVT model B~\cite{Pappadopulo:2014qza}.

\begin{figure*}[htbp]
\centering
\includegraphics[width=0.3\textwidth]{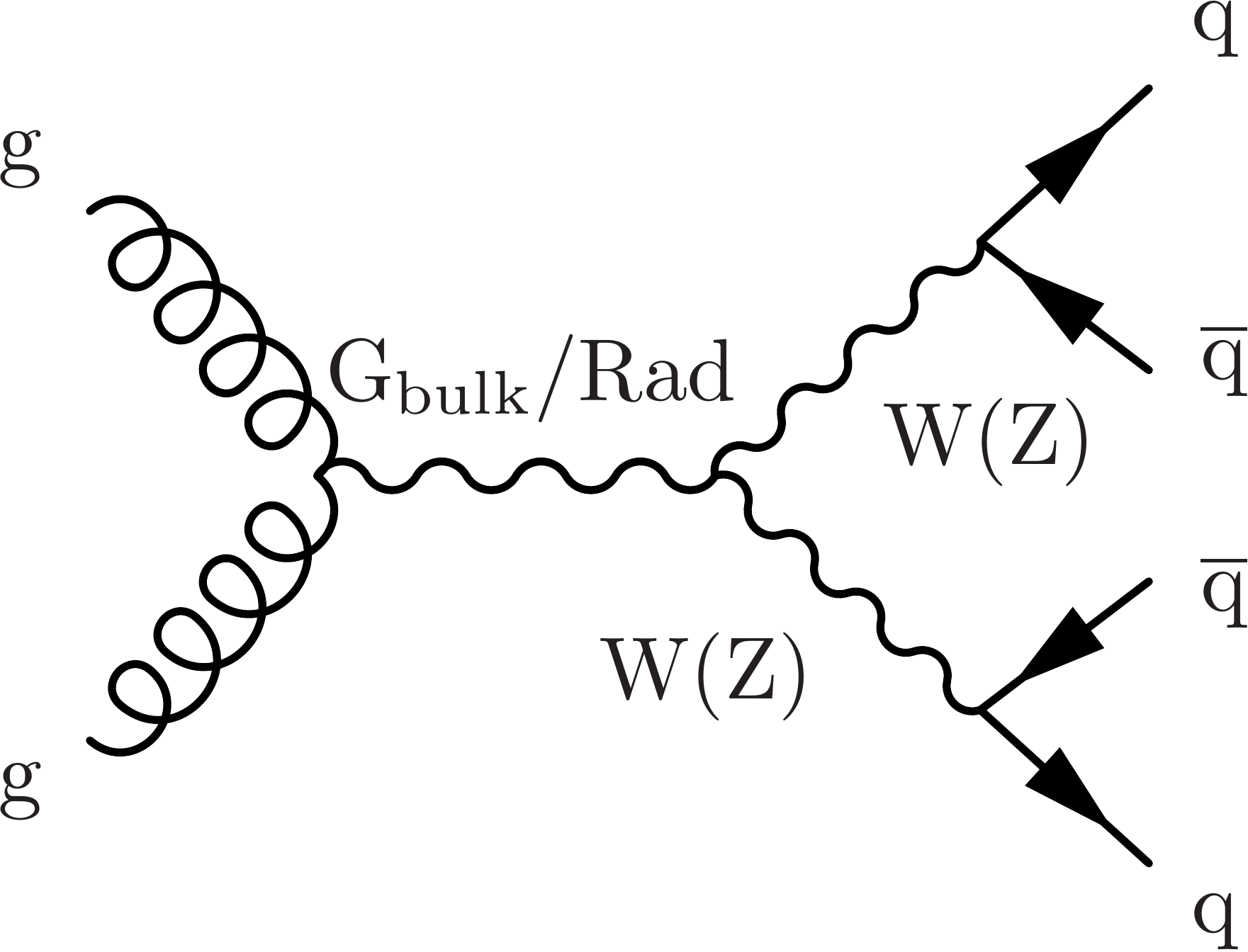}
\hfill
\includegraphics[width=0.3\textwidth]{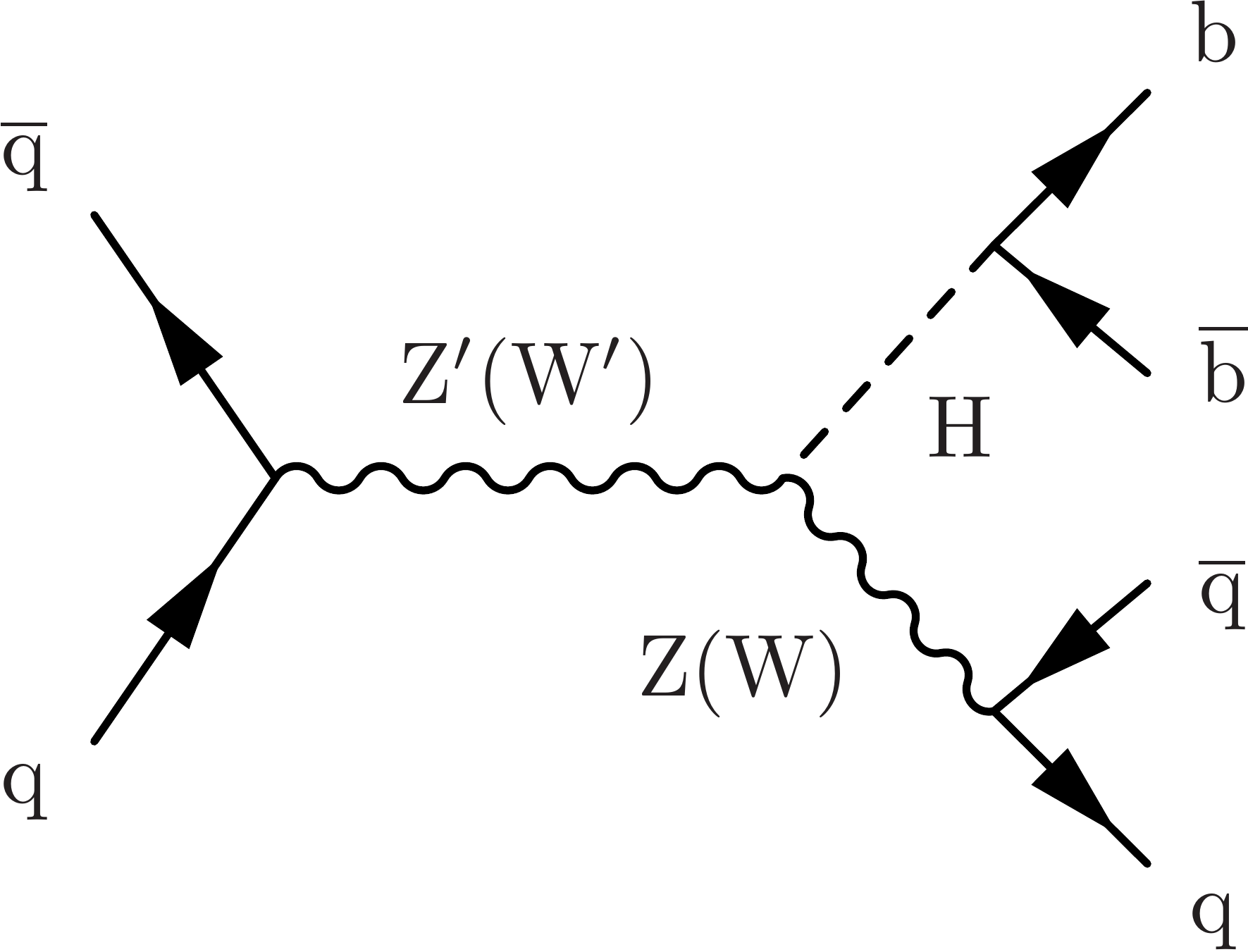}
\hfill
\includegraphics[width=0.3\textwidth]{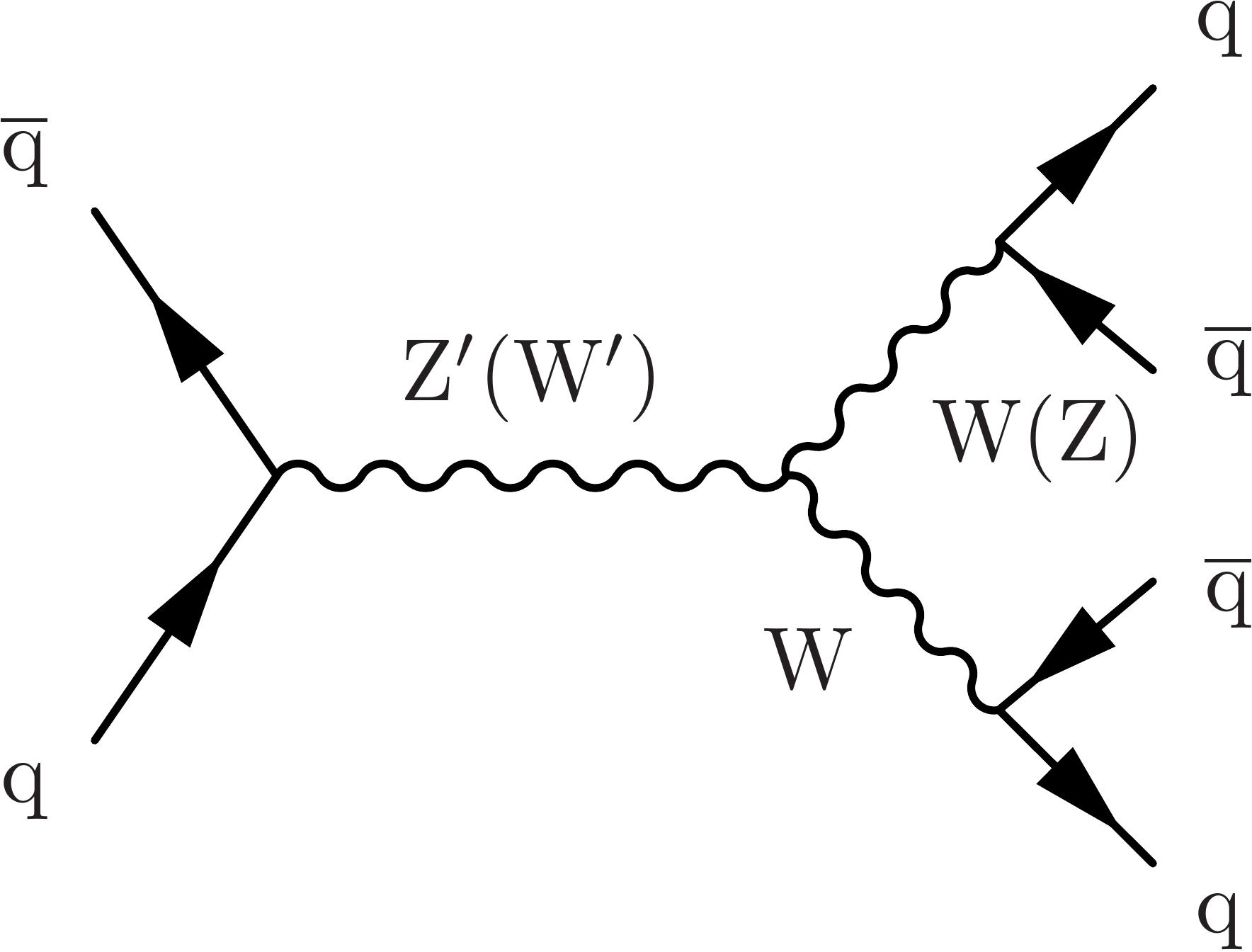}\\
\includegraphics[width=0.3\textwidth]{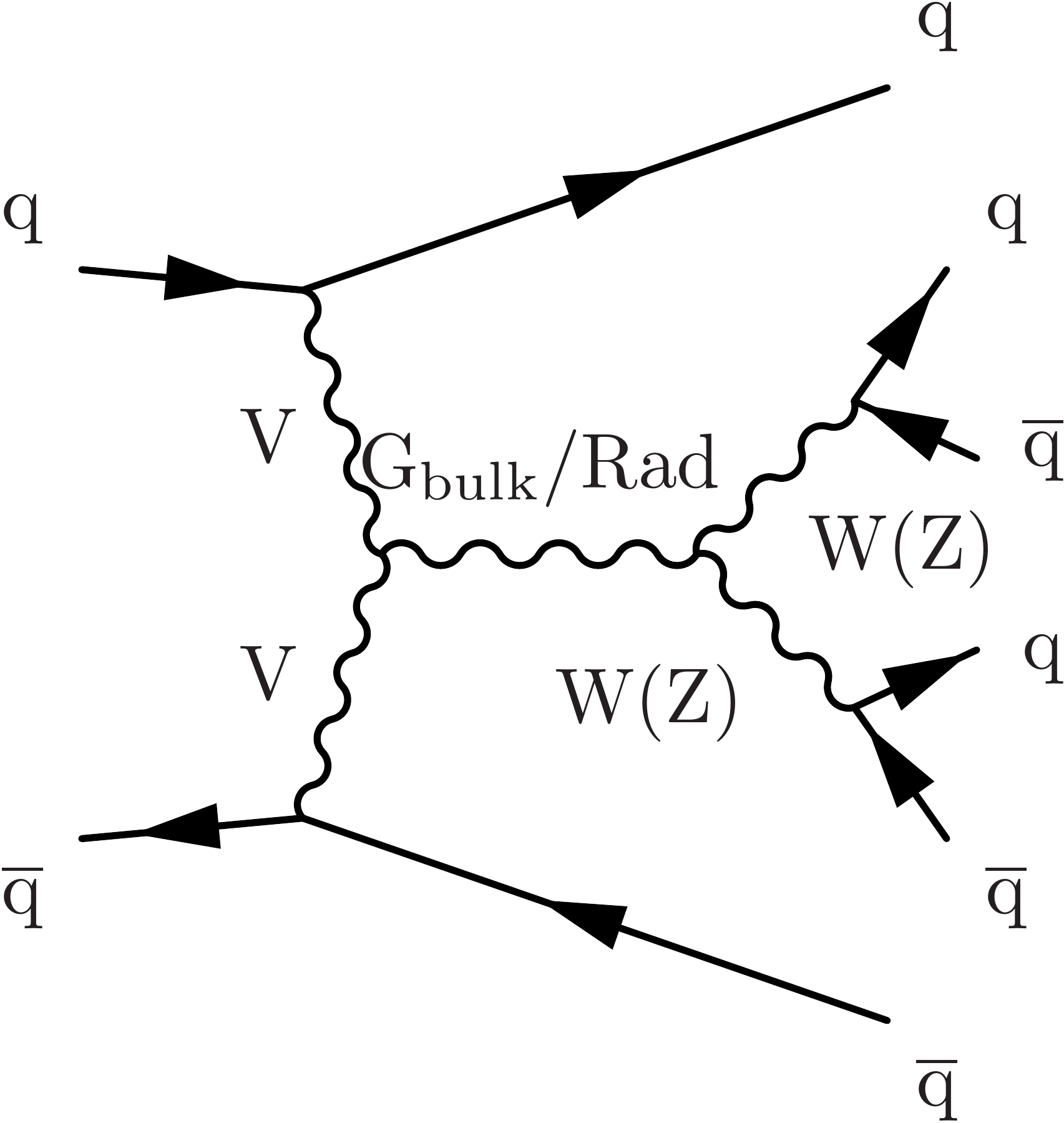}
\hfill
\includegraphics[width=0.3\textwidth]{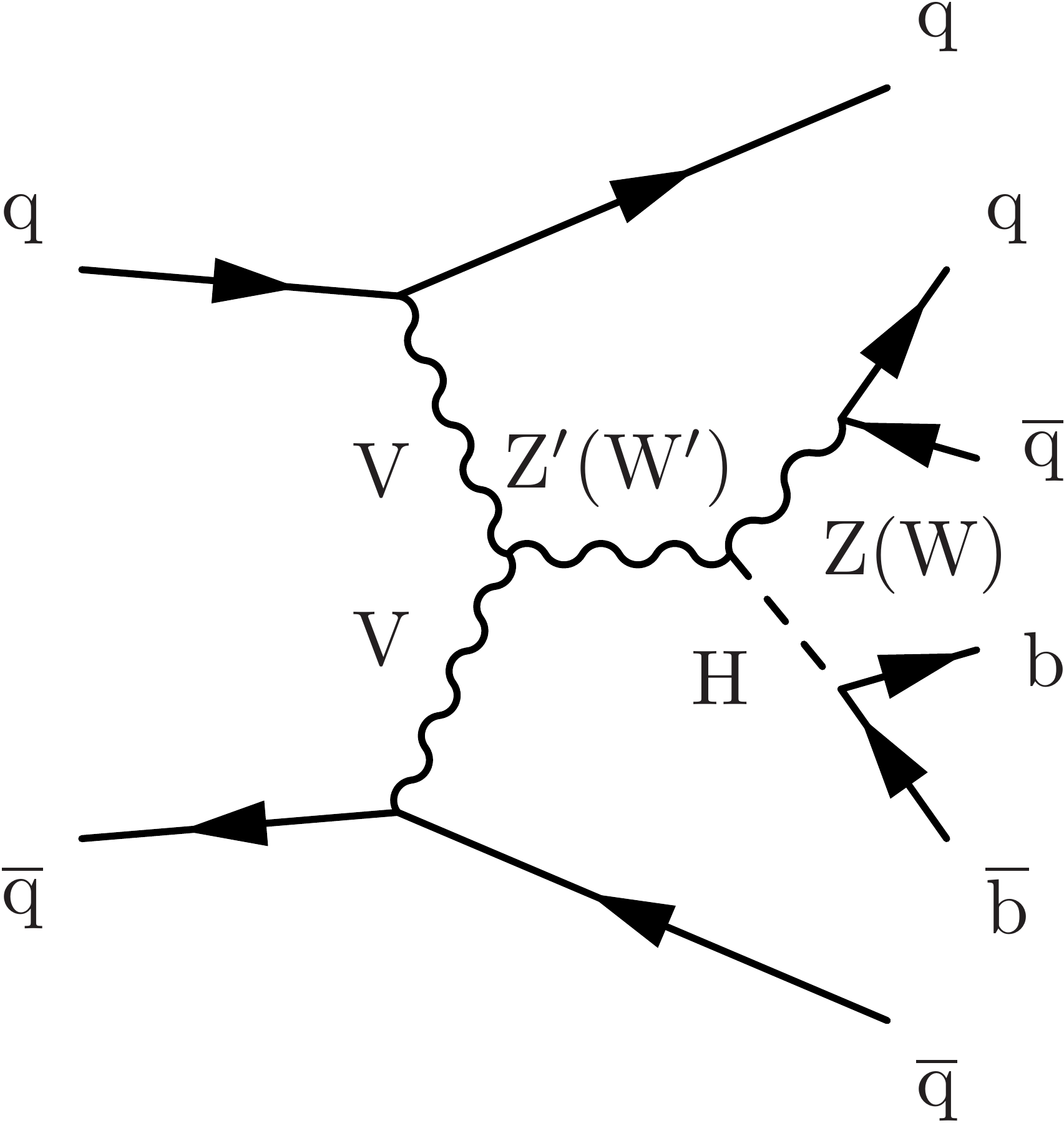}
\hfill
\includegraphics[width=0.3\textwidth]{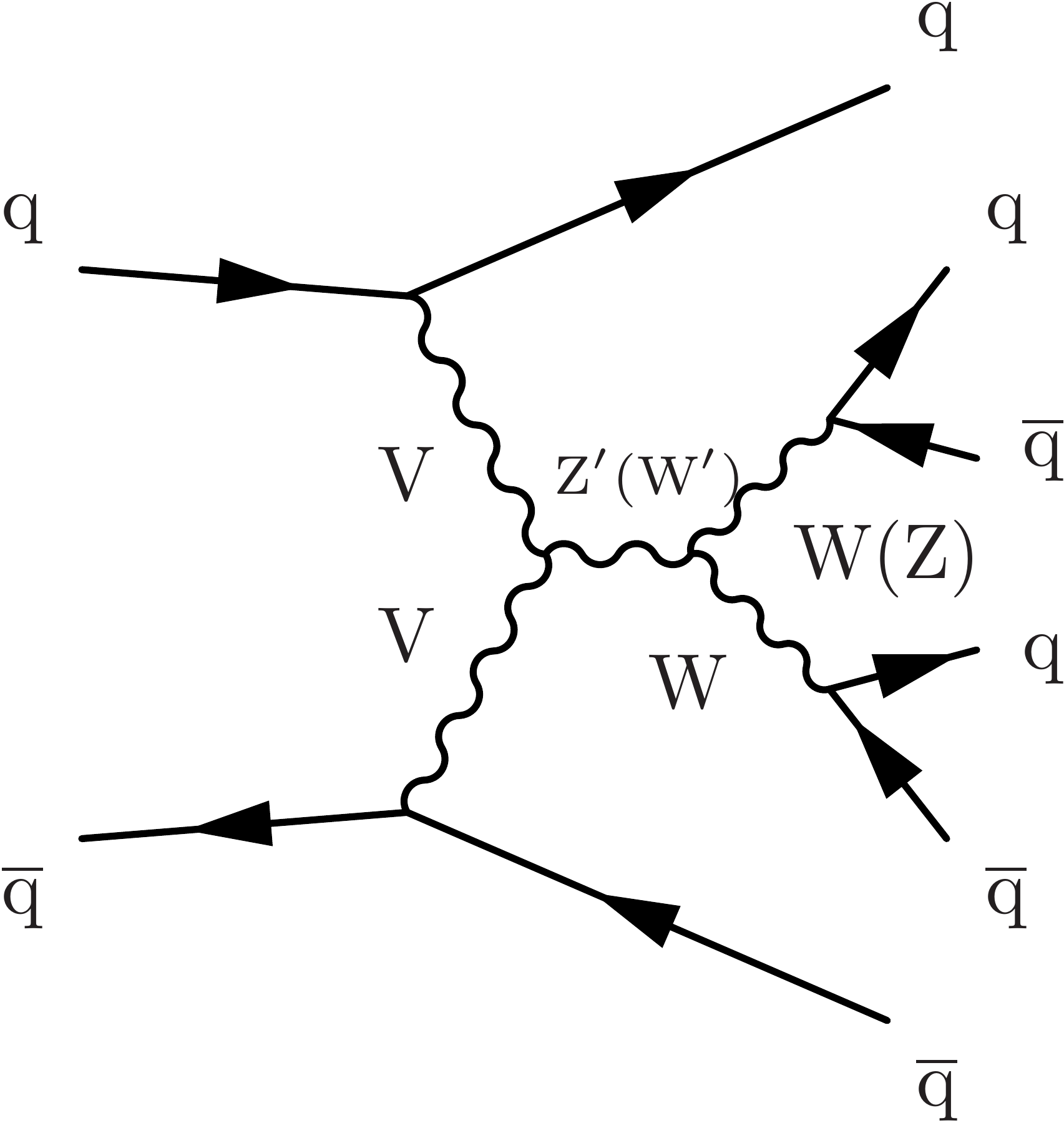}

\caption{Feynman diagrams of the signal processes ggF or DY produced (upper) and VBF produced (lower): 
(left) graviton or radion decaying to \WW or \ZZ;
(center) \PZpr and \PWpr decaying to \ZH and \WH, respectively;
(right) \PZpr and \PWpr decaying to \WW and \WZ, respectively.}
\label{fig:feynman}
\end{figure*}

This Letter presents a search in the all-jets final state for new heavy resonances using a \pp collision data set collected by the CMS experiment at a centre-of-mass energy of 13\TeV in 2016--2018 corresponding to an integrated luminosity of \INTLUMI.
Resonances decaying to a \VV or \VH boson pair with masses between \BulkGMassMinNoTeV and \BulkGMassMax and produced via Drell--Yan (DY), gluon fusion (ggF), or vector boson fusion (VBF) are targeted.
Representative Feynman diagrams for the aforementioned processes are shown in Fig.~\ref{fig:feynman}.

Because of the large Lorentz boost of the \PH, \PW, and \PZ bosons from the resonance decay, each boson decay is typically clustered as a single large-radius jet.
The final state thus consists of two large-radius jets (distance parameter $R=0.8$) in the case of DY and ggF production, with two additional small-radius ($R=0.4$) jets in the case of VBF production.
The SM background estimation and signal extraction procedure is based on a three-dimensional (3D) maximum likelihood fit to the mass of the two large-radius jet systems and the two individual large-radius jet masses, as introduced in a previous search by the CMS Collaboration in the \VV channel~\cite{Sirunyan:2019jbg}.
The sensitivity to \VV and \VH resonances is significantly improved compared to previous searches by categorizing events according to jet tagging algorithms based on machine learning~\cite{Sirunyan:2020lcu} that analyze the substructure of the large-radius jets to separate jets that originate from boosted \PH, \PW, and \PZ bosons from other jets.
Events are further categorized based on the presence of additional small-radius jets, enhancing the sensitivity to VBF-produced resonances.

\section{The CMS detector and event reconstruction}
\label{sec:cmsdetector}

The central feature of the CMS apparatus is a superconducting solenoid of 6\unit{m} internal diameter, providing a magnetic field of 3.8\unit{T}. Within the solenoid volume are a silicon pixel and strip tracker, a lead tungstate crystal electromagnetic calorimeter, and a brass and scintillator hadron calorimeter, each composed of a barrel and two endcap sections. Forward calorimeters extend the pseudorapidity coverage provided by the barrel and endcap detectors. Muons are measured in gas-ionization detectors embedded in the steel flux-return yoke outside the solenoid. A more detailed description of the CMS detector, together with a definition of the coordinate system used and the relevant kinematic variables, can be found in Ref.~\cite{Chatrchyan:2008zzk}.

Event reconstruction is based on a particle flow algorithm~\cite{Sirunyan:2017ulk}, which reconstructs and identifies individual particles (photon, electron, muon, charged hadron, neutral hadron) with information from the various elements of the CMS detector.
Jets are reconstructed from these particles, using the anti-\kt jet clustering algorithm~\cite{Cacciari:2008gp} with distance parameters of $R=0.4$ (AK4 jets) and $R=0.8$ (AK8 jets), as implemented in the \FASTJET package~\cite{Cacciari:2011ma}.
To mitigate the effect of additional \pp interactions within the same or nearby bunch crossings (pileup) on the reconstructed jet momentum,
tracks identified as originating from pileup vertices are discarded and an offset correction is applied to correct for remaining contributions.
Jet energy corrections are derived from simulation studies. In situ measurements of the momentum balance in dijet, photon+jet, \Zjets, and multijet events are used to determine any residual differences between the jet energy scale in data and in simulation, and appropriate corrections are made~\cite{Khachatryan:2016kdb}. Additional selection criteria are applied to each jet to remove jet measurements potentially arising from instrumental effects or reconstruction failures~\cite{CMS-PAS-JME-16-003}.

Events of interest are selected using a two-tiered trigger system~\cite{Khachatryan:2016bia}. The first level, composed of custom hardware processors, uses information from the calorimeters and muon detectors to select events at a rate of around 100\unit{kHz} within a fixed time interval of less than 4\mus. The second level, known as the high-level trigger, consists of a farm of processors running a version of the full event reconstruction software optimized for fast processing, and reduces the event rate to around 1\unit{kHz} before data storage.
Events are selected online with a variety of different jet triggers based on the highest jet transverse momentum (\PT) or the \PT sum of all jets in the event (\HT). For some of these triggers additional requirements on the trimmed jet mass~\cite{Krohn:2009th} are applied to allow lower the \PT and \HT thresholds~\cite{Sirunyan:2019jbg,Sirunyan:2017acf}.
The trigger efficiency as a function of the invariant mass of the two highest \PT AK8 jets (\MVV) is $>$99\% above \Mthr for all three data-taking years, and the subsequent analysis thus requires \MVV to be above this threshold.

\section{Signal and background simulation}
\label{sec:samples}

Each signal model is characterized by key parameters.
The bulk graviton model is characterized by two free parameters: the mass of the first Kaluza--Klein excitation of a spin-2 boson (the Kaluza--Klein bulk graviton), and the ratio $\kappatilde=\kappa\sqrt{8\pi}/\Mpl$, with $\kappa$ being the unknown curvature scale of the extra dimension and \Mpl the Planck mass. A scenario with $\kappatilde=0.5$, resulting in resonances with a width smaller than the detector resolution is considered in this analysis, as motivated in Ref.~\cite{Oliveira:2014kla}.
The radion model is also characterized by two parameters: \rc, the compactification radius, and \lambdaR, the ultraviolet cutoff of the theory. The scenario with $\kappa \rc \pi = 35$ and $\lambdaR = 3\TeV$~\cite{Oliveira:2014kla} is considered in this analysis.
The HVT model is characterized in terms of four parameters: the mass of the \PWpr and \PZpr resonance; a coefficient \cF, which scales the couplings of the additional gauge bosons to fermions; \cH, which scales the couplings to the Higgs boson and longitudinally polarized SM vector bosons; and \gV, representing the typical strength of the new vector boson interaction.
Two scenarios are considered in this analysis: HVT model B, corresponding to $\gV=3$, $\cH=-0.98$, and $\cF=1.02$~\cite{Pappadopulo:2014qza}; and HVT model C~\cite{Pappadopulo:2014qza}, corresponding to $\gV=1$, $\cH=1-3$, and $\cF=0$. In both scenarios, the new resonances have a narrow decay width and large branching fraction to vector boson pairs, while the fermionic couplings are suppressed. In the HVT model C, which has no fermionic couplings, the resonances would be produced at the LHC exclusively via the VBF mode.

Monte Carlo (MC) simulated events of the radion, bulk graviton, and HVT resonance signal processes are generated at leading order (LO) in perturbative quantum chromodynamics (QCD) with \MGvATNLO versions
2.4.2 and 2.6.0~\cite{Alwall:2014hca}.
The parton showering and hadronization is simulated with \PYTHIA versions 8.205 and 8.230~\cite{Sjostrand:2014zea}, for 2016 and 2017--2018 detector conditions, respectively.
The NNPDF 3.0~\cite{Ball:2014uwa} LO parton distribution functions (PDFs) are used together with the CUETP8M1~\cite{Khachatryan:2015pea} and CP5~\cite{CMS:2019csb} underlying-event tunes in \PYTHIA for 2016 and 2017--2018 conditions, respectively.
The signal cross sections are computed at next-to-LO (NLO) with \MGvATNLO with the PDF4LHC15\_100 PDF set~\cite{Butterworth:2015oua, Dulat:2015mca, Harland-Lang:2014zoa, Ball:2014uwa, Gao:2013bia, Carrazza:2015aoa}.

Simulated event samples of the SM background processes are used to develop the analysis strategy and create templates for distributions used in the comparison with data.
The QCD multijet production is simulated with three generator configurations: \PYTHIA only, the LO mode of \MGvATNLO~\cite{Alwall:2007fs} interfaced with \PYTHIA for the parton shower evolution and matching (\textsc{mg}+\PYTHIA{8} in the following), and \HERWIGpp 2.7.1~\cite{Bahr:2008pv} with the CUETHS1 tune~\cite{Khachatryan:2015pea}.
Top quark pair (\ttbar), single top quark, and boson pair production are modelled at NLO with \POWHEG v2~\cite{Alioli:2011as} interfaced with \PYTHIA.
The production of \Wjets and \Zjets (\Vjets) is simulated at LO with \MGvATNLO interfaced with \PYTHIA. The same underlying-event tunes as used in the signal event samples are used in the background event samples. A correction~\cite{Kallweit:2014xda} is applied to the simulated \Vjets events to match the \PT distribution of the vector bosons computed at LO in QCD to the one predicted at NLO in QCD, and another correction~\cite{Kallweit:2015dum} is used to account for NLO electroweak effects at high \PT. The NNPDF 3.1~\cite{Ball:2017nwa} next-to-NLO (NNLO) PDFs are employed for simulated \Vjets events.

All samples are processed through a \GEANTfour-based~\cite{Agostinelli:2002hh} simulation of the CMS detector. To simulate the effect of pileup collisions, additional inelastic events are generated using \PYTHIA and superimposed on the hard-scattering events. The simulated events are weighted to reproduce the distribution of the number of reconstructed pileup vertices observed in the 2016, 2017, and 2018 data separately.
While the detector components and conditions varied across the three years of data taking, the detector performance relevant to this analysis, in particular with regard to the \MVV and \MJ scale and resolution, was very similar~\cite{CMS-DP-2020-019,CMS-DP-2021-033}.
We therefore combine the simulated event samples corresponding to the three years of data taking, weighting them according to the corresponding integrated luminosity
of \INTLUMIsix,  \INTLUMIseven, and \INTLUMIeight\fbinv for 2016, 2017, and 2018~\cite{CMS:2021xjt,CMS-PAS-LUM-17-004,CMS-PAS-LUM-18-002}, respectively.

\section{Event selection}
\label{sec:eventreconstruction}

All events are required to have at least one primary vertex reconstructed within a region extending 24\cm along the beam, centred on the centre of the detector, and extending 2\cm from the mean beam axis in the transverse direction~\cite{Chatrchyan:2014fea}.
The primary vertex is taken to be the vertex corresponding to the hardest scattering in the event, evaluated using tracking information alone, as described in Section 9.4.1 of Ref.~\cite{CMS-TDR-15-02}.

Events are selected by requiring at least two AK8 jets with $\PT>200\GeV$ and $\abs{\eta}<2.5$. The two AK8 jets with the highest \PT in the event are selected as potential vector boson or Higgs boson candidates and are required to have a separation of $\abs{\Delta\eta^{\text{AK8}}}<1.3$ in order to reduce the QCD multijet background.
The invariant mass of the two AK8 jets is required to fulfill $\MVV>\Mthr$, based on the trigger selection discussed in Section~\ref{sec:cmsdetector}.
Given the \MVV resolution of about 10\%, the lowest resonance mass that passes this selection with high efficiency is \BulkGMassMin.
Jets originating from the misreconstruction of a high-momentum lepton are rejected by requiring an angular separation $\DR = \sqrt{\smash[b]{(\Delta\eta)^2 + (\Delta\phi^2)}} >0.8$ (where $\phi$ is the azimuthal angle) from muons (electrons) with $\PT > 20$ (35)\GeV and satisfying identification criteria optimized for high-momentum leptons~\cite{Sirunyan:2018fpa,Khachatryan:2015hwa}.

Hadronic decays of \PH, \PW, and \PZ bosons are identified using two variables computed from the AK8 jet constituents: the ``groomed'' mass of the jet and the score from the DeepAK8 neural network~\cite{Sirunyan:2020lcu} jet tagging algorithm.
When computing these jet variables,
the pileup per particle identification algorithm~\cite{Sirunyan:2020foa,Bertolini:2014bba} is used to mitigate the effect of pileup at the reconstructed particle level, making use of local shape information, event pileup properties, and tracking information.

To improve the resolution of the jet mass, the groomed jet mass (\MJ) is calculated with the \emph{soft drop} algorithm~\cite{Larkoski:2014wba}, which removes soft and wide angular radiation
by reclustering the jet constituents using the Cambridge--Aachen algorithm~\cite{Dokshitzer:1997in,Wobisch:1998wt}
with parameters $\beta=0$, $z_\text{cut}=0.1$, and $R_0 = 0.8$.

The DeepAK8 tagger is trained on multiple properties of the reconstructed jet constituents and secondary vertices. For each jet constituent, measured properties such as \PT, charge, and angular separation between the particle and the jet axis, are included. Additional properties of charged particles and secondary vertices are used to extract features related to the presence of heavy-flavor (b or c) quarks. The DeepAK8 tagger features a two-step customized deep neural network architecture, developed to handle the high number of input features and their correlations. In the first step, two separate convolutional neural networks extract the helpful characteristics of the constituent particles and the secondary vertices. The jet categorization happens in the second step, where a simple, fully connected network combines and processes the output of the two convolutional networks.

Jets are categorized according to their origin from single quarks or gluons, or \PH, \PW, and \PZ bosons decaying to bottom, charm, or light quark pairs.
Two tagger discriminants based on the DeepAK8 categorization are used in this search.
The \DeepWZqq tagger considers \PW and \PZ bosons decaying to light quark pairs as signal and single quarks and gluons as background.
The \DeepZHbb tagger considers \PH and \PZ bosons decaying to bottom quark pairs as signal and single quarks and gluons as background.
As the background estimation procedure of this search relies on a smooth shape for the groomed jet mass after the DeepAK8 tagger selection criteria are applied, an adversarial training of the DeepAK8 tagger~\cite{Sirunyan:2020lcu} with a reduced correlation between groomed jet mass and neural network output is used, and an additional decorrelation procedure~\cite{Sirunyan:2020lcu,ddt} is applied.
The selection on the neural network output is chosen such that it yields a constant tagging rate as a function of \PT and groomed jet mass for the quark or gluon jet with the highest \PT in \MGvATNLO QCD multijet simulation.
The taggers are labeled according to their expected average misidentification rate in percent for quark and gluon jets estimated with the \MGvATNLO QCD multijet simulation.
The groomed jet mass, the \DeepWZqq tagging efficiency, and the probability of the \DeepWZqq tagger to misidentify hadronically decaying top quarks are calibrated in a \ttbar event sample enriched in hadronically decaying \PW bosons~\cite{Sirunyan:2020lcu}.
The \DeepZHbb tagger is calibrated in a sample enriched with $\Pg\to\bbbar$ jets using a double-muon tag~\cite{Sirunyan:2017ezt}.

To reduce the QCD multijet background, we require each AK8 jet mass to be between 55 and 215\GeV.
Finally, a loose requirement of $\rho=\ln((\MJ)^2/\PT^2) < -1.8$~\cite{ddt} is applied in order to veto events in which the jet mass is high, but the jet \PT is low.
In those cases, the cone size of $\DR=0.8$ is too small to contain the full jet, affecting both the jet mass resolution and the \DeepWZqq tagging efficiency, which are consequently not well modelled in the simulation.

To simplify the modelling of the 3D shapes in the (\MVV, \MJO, \MJT) space, the two AK8 jets are labelled at random so that the mass distributions of the first and second selected jet, \MJO and \MJT, have the same shape.
The selected events are subdivided to enhance sensitivity to VBF signal production.
The VBF topology is selected by requiring at least two AK4 jets with $\PT>30\GeV$ and $\abs{\eta}<5.0$ that do not overlap within $\DR<1.2$ with the leading two AK8 jets.
A separation of $\abs{\Delta\eta^{\text{AK4}}}>4.5$ and an invariant mass of the two leading AK4 jets larger than 800\GeV is required to maximize the signal-to-background ratio and achieve the best sensitivity.

\begin{table*}[tb]
  \centering\renewcommand\arraystretch{1.2}
  \topcaption{Summary of the event category definitions. The categories are listed from the highest to the lowest sensitivity in each production mode. The percentages correspond to the maximum misidentification rate associated with the high (HP) and low (LP) purity working points.}
  \begin{tabular}{lccc}
    Event category & VBF & AK8 jet\textsubscript{1 or 2} tag & AK8 jet\textsubscript{2 or 1} tag \\
    \hline
    VBF \VH HPHP & pass & (5\% \DeepWZqq or 2\% \DeepZHbb) & 2\% \DeepZHbb\\
    VBF \VV HPHP & pass & 5\% \DeepWZqq & 5\% \DeepWZqq\\
    VBF \VH LPHP & pass & (20\% \DeepWZqq or 10\% \DeepZHbb) & 2\% \DeepZHbb\\
    VBF \VH HPLP & pass & 5\% \DeepWZqq & 10\% \DeepZHbb\\
    VBF \VV HPLP & pass & 5\% \DeepWZqq & 20\% \DeepWZqq\\
    DY/ggF \VH HPHP  & fail & (5\% \DeepWZqq or 2\% \DeepZHbb) & 2\% \DeepZHbb\\
    DY/ggF \VV HPHP & fail & 5\% \DeepWZqq & 5\% \DeepWZqq\\
    DY/ggF \VH LPHP & fail & (20\% \DeepWZqq or 10\% \DeepZHbb) & 2\% \DeepZHbb\\
    DY/ggF \VH HPLP & fail & 5\% \DeepWZqq & 10\% \DeepZHbb\\
    DY/ggF \VV HPLP & fail & 5\% \DeepWZqq & 20\% \DeepWZqq\\
  \end{tabular}
  \label{tab:categories}
\end{table*}

Events are further grouped into categories using the \DeepWZqq and \DeepZHbb taggers with different tagging rates, \ie different signal-to-background ratios (high purity ``HP'' and low purity ``LP''), to target specific signal decay modes, as summarized in Table~\ref{tab:categories}. Purities are expressed in terms of the maximum allowed misidentification rate.
To ensure that each event ends up only in one category, events are required to fail the selection of all previous categories in the order listed in Table~\ref{tab:categories}.
The overall signal efficiency depends on the resonance type and mass, and is estimated from simulation to be 12--26\% and 2--18\% for DY/ggF production and VBF production, respectively.
For resonances decaying to a Higgs boson, the \VH HPHP category contains 40--60\% of the signal, while for the other resonances 10--40\% of the signal is contained in the \VH HPHP category and 15--40\% in the \VV HPHP category. The other DY/ggF categories contain 5--40\% of the signal, depending on the resonance type and mass.
The VBF categories contain 25--40\% of the VBF signals and less than 5\% of the DY/ggF signals.

\section{Background estimation and signal extraction}
\label{sec:analysis}

To test for the presence of narrow resonances decaying to two bosons, a 3D maximum likelihood fit of signal and background templates to the data in the (\MVV, \MJO, \MJT) space is carried out in all event categories simultaneously.
We give a brief explanation of the fit model in the following, and refer to Ref.~\cite{Sirunyan:2019jbg} for more details.

The main SM background processes in this search are QCD multijet, \ttbar, \Wjets, and \Zjets production.
The QCD multijet production is the largest background in all categories.
While in the \VV event categories the QCD multijet production is by far the most dominant, making up more than 75\% of the background, in the \VH categories \ttbar contributes up to 41\% of the background.
The background from \Wjets and \Zjets contributes up to 4\%, most significantly to the LP categories.
The background contribution from single top quark and diboson production is less than 1.5 and 0.5\%, respectively, and is modelled as a part of the \ttbar background.
Signal and background templates in the (\MVV, \MJO, \MJT) space are built from simulation with the method described in Ref.~\cite{Sirunyan:2019jbg} to prevent fluctuations due to the limited number of generated events.
While the signal is resonant in the \MVV, \MJO, and \MJT observables, all backgrounds are nonresonant in \MVV.
The QCD multijet background is nonresonant in all three dimensions.
The remaining backgrounds have partly resonant components in \MJ, modelled by separate templates, and partly nonresonant components that are absorbed in the QCD multijet background template.
In the background model it is assumed that background components resonant in \MJ containing jets from genuine \PV bosons or top quarks can be reasonably well modelled by simulation calibrated with external measurements. Nonresonant background components with jets initiated by single quarks or gluons misidentified by the \DeepWZqq and \DeepZHbb taggers may be largely mismodelled in simulation, and are constrained in the likelihood fit to the data.
Systematic uncertainties in the signal and background models are treated as nuisance parameters and profiled in the statistical interpretation of the data.

\begin{figure*}[htbp]
\centering
\includegraphics[width=0.49\textwidth]{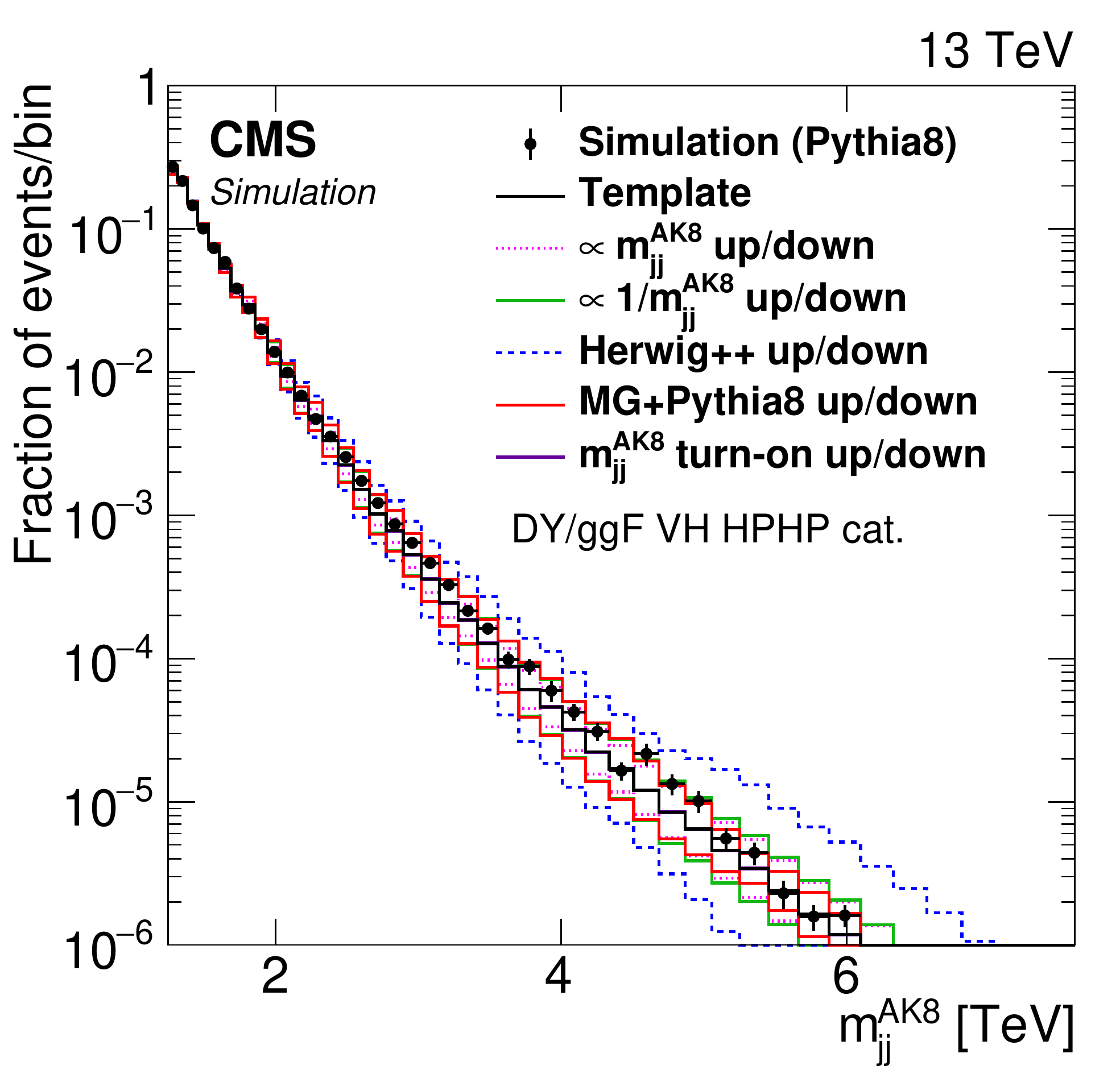}
\includegraphics[width=0.49\textwidth]{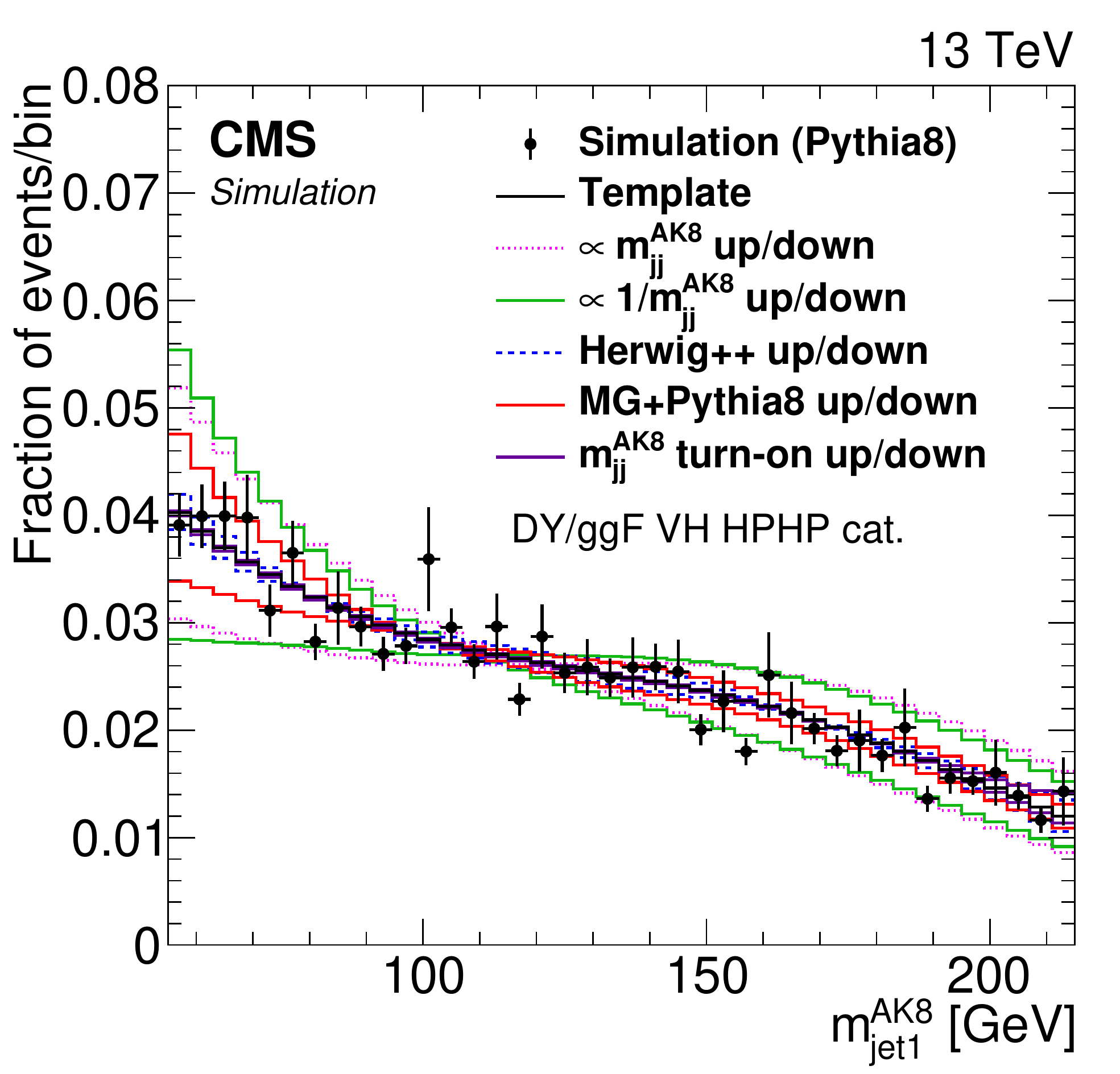}
\caption{Distributions of \MVV (left) and \MJO (right) in the nominal QCD multijet simulation using \PYTHIA{8} (black markers) and three-dimensional templates (black solid line) in the DY/ggF \VH HPHP category. Superimposed are the normalized up and down variations of five alternative shapes corresponding to shape nuisance parameters in the fit model. Each of the variations has a characteristic dependence on \MJ to allow the necessary flexibility in the fit for the template to adapt to the data. We also allow an overall variation of the rate in the fit. The distribution of \MJT is not shown, as it is similar to the distribution of \MJO.}
\label{fig:sys_QCD}
\end{figure*}

To account for discrepancies in the QCD multijet simulation and data, we allow the background model (template) to adapt to the data using physically motivated variations of the nominal shape of the distributions obtained from the \PYTHIA{8}-only simulations. The templates are smooth versions of the simulations  obtained from simulated events through a forward folding approach, as described in Ref.~\cite{Sirunyan:2019jbg}. The alternative shapes described below are shown in Fig.~\ref{fig:sys_QCD} together with the \PYTHIA{8} templates and simulation.
The normalization of the QCD multijet background is allowed to vary within 50\% and is treated as uncorrelated between all event categories to let the fit constrain the yield in each of them separately.
Alternative shapes with variations proportional to \MVV, 1/\MVV, \MJ, and 1/\MJ allow for variations in the underlying jet \PT spectrum and jet mass scale.
Two additional alternative shapes that simultaneously affect the resonance mass and the groomed jet mass are also added in order to take into account differences in the MC generation and modelling of the parton shower.
These alternative templates are derived using the \HERWIGpp and \textsc{mg}+\PYTHIA{8} QCD multijet simulation.

For events with a large \MJ ($>$175\GeV) and low \MVV($<$1400\GeV), the distribution exhibits a turn-on close to the $\rho<-1.8$ threshold that is particularly difficult to model with the QCD multijet simulation. An additional smooth shape variation (\MVV turn-on in Fig.~\ref{fig:sys_QCD}) parameterizing any discrepancy between the 3D template and the QCD multijet simulation in this region is added to the fit.
We neglect the uncertainties due to the choice of PDF in the QCD multijet simulation, since it was verified that these effects are covered by the included shape variations.
All shape variations are treated as uncorrelated among the event categories with different tagger requirements and correlated among the VBF and DY/ggF categories.

Similarly, for the \Wjets and \Zjets backgrounds, two alternative shapes of the \MVV templates derived by a variation proportional to \MVV and $1/\MVV$ are considered, and the normalization is allowed to vary within 50\%, uncorrelated between all event categories. We neglect the uncertainties due to the NLO QCD+EW corrections to the \Vjets simulation, since it was verified that these effects are covered by the included shape variations.
The \ttbar background normalization is allowed to vary within 6\%~\cite{Sirunyan:2647715,Khachatryan:2224321} and is considered correlated between the event categories as it is dominantly doubly-resonant in \MJO and \MJT.
An alternative shape where the top quark \PT spectrum is reweighted to match the observed \PT distribution in \ttbar data~\cite{Khachatryan:2224321} is also taken into account.

Systematic uncertainties in \PH, \PW, and \PZ boson reconstruction and identification are dominant for the signal, and are treated as correlated between the signal and the background components resonant in the \MJ observable.
The 3D fit constrains the corresponding nuisance parameters that are included with a prior uncertainty from external measurements of the \PH, \PW, and \PZ boson reconstruction and identification performance.
Uncertainties in the jet mass scale, jet mass resolution, and the \DeepWZqq and \DeepZHbb tagging efficiencies are estimated from their calibration in a semileptonic \ttbar event sample~\cite{Sirunyan:2020lcu} and in a sample enriched with $\Pg\to\bbbar$ jets using a double-muon tag~\cite{Sirunyan:2017ezt}.
The jet mass scale and resolution are allowed to vary within 2 and 8\%, respectively.
The tagging efficiency of the 5\% \DeepWZqq, 20\% \DeepWZqq, 2\% \DeepZHbb, and 10\% \DeepZHbb taggers are allowed to vary within 7--15\%, 7--9\%, 3--9\%, and 2--7\%, respectively, depending on the year of data taking.
Uncertainties equal to 1.5 times the nominal tagging uncertainty values are included in the fit as additional nuisance parameters to account for the potential mismodelling of the \PT dependence of the tagging efficiency beyond the range of the data-to-simulation scale factor measurement.
Finally, the \DeepWZqq tagger's rate for misidentification of hadronically decaying top quarks in the \ttbar sample is allowed to vary within 3--7\%.

The uncertainty in the signal acceptance and in the mean and width of the \MVV, \MJO, and \MJT distributions originating from the PDFs for the signal processes are determined from the LO NNPDF set, where the root-mean-square of 100 pseudo-experiments obtained from the PDF set provides the uncertainty envelope.
Uncertainties related to the jet energy scale and resolution are taken into account as nuisance parameters varying the \MVV mean and width.
The uncertainty in the integrated luminosity amounts to 1.6\%~\cite{CMS:2021xjt,CMS-PAS-LUM-18-002,CMS-PAS-LUM-17-004}.

\begin{figure*}[!htbp]
\centering
\includegraphics[width=0.49\textwidth]{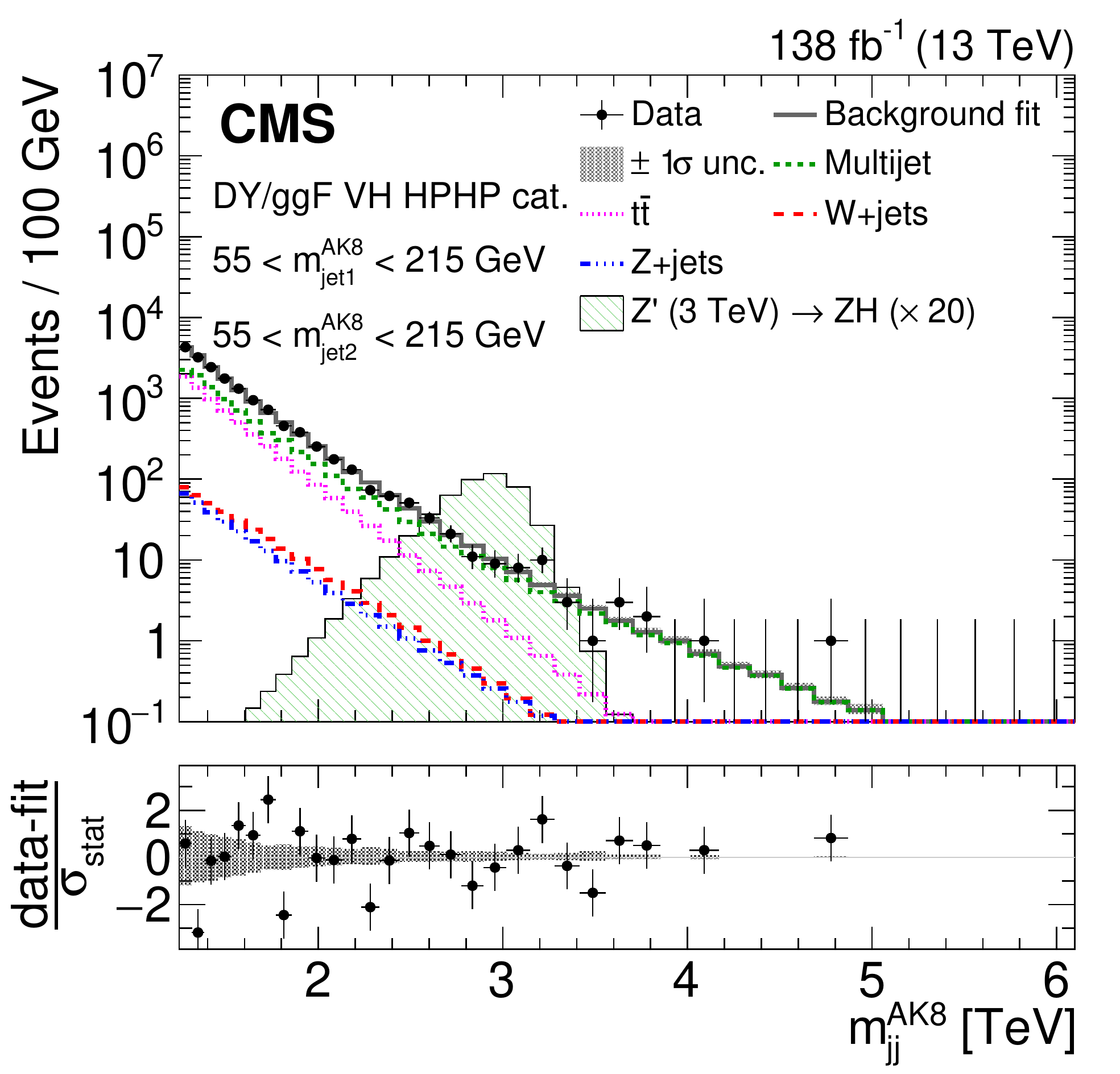}
\includegraphics[width=0.49\textwidth]{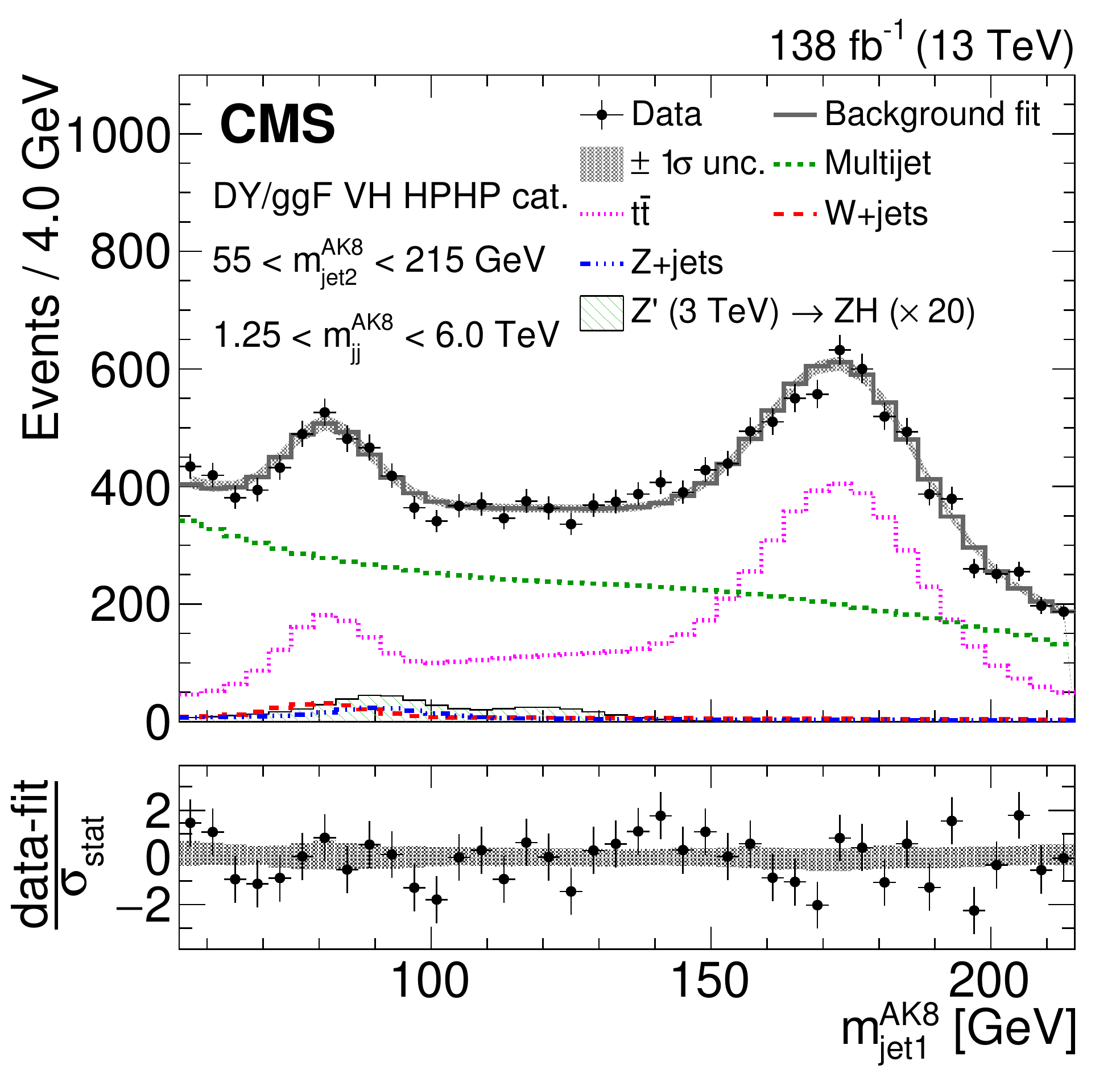}
\caption{Comparison between the background post-fit and the data distributions of \MVV (left) and \MJO (right) in the DY/ggF \VH HPHP category.
The background shape uncertainty is shown as a gray shaded band around the result of the maximum likelihood fit to the data under the background-only assumption (gray solid line), and the statistical uncertainties in the data are represented as vertical bars. The various background components contributing to the total background fit are also shown with different line colors.
An example of a signal distribution is overlaid, where the number of expected events is scaled by an arbitrary normalization factor of 20.
Shown below each mass plot is the difference between the data and the fit divided by the statistical uncertainty in the data; the uncertainty bar represents the statistical uncertainty only. The total uncertainty in the background estimate fitted to the data divided by the statistical uncertainty in the data is shown as a band.}
\label{fig:postfit1}
\end{figure*}

Figure~\ref{fig:postfit1} shows the \MJ and \MVV spectra in data in the category most sensitive to the DY/ggF \VH signal.
The solid gray histograms represent the results of the maximum likelihood fit to the data under the background-only assumption.
The resonant background components are shown separately.
A signal is superimposed onto all projections corresponding to a signal yield as expected from the theoretical prediction and the analysis selection efficiency, and scaled by an arbitrary factor.
The lower panels in Fig.~\ref{fig:postfit1} show the difference between the data and the fit divided by the statistical uncertainty in the data. It should be noted that the fit uncertainty and the statistical uncertainty are partially correlated. Thus, it is not ensured that the plotted quantity follows a Gaussian distribution~\cite{CDF:AN5776} and the Gaussian assumption becomes even less valid in bins with a low number of events, \ie in the last few bins in each distribution.

\begin{figure*}[t!]
\centering
\includegraphics[width=0.49\textwidth]{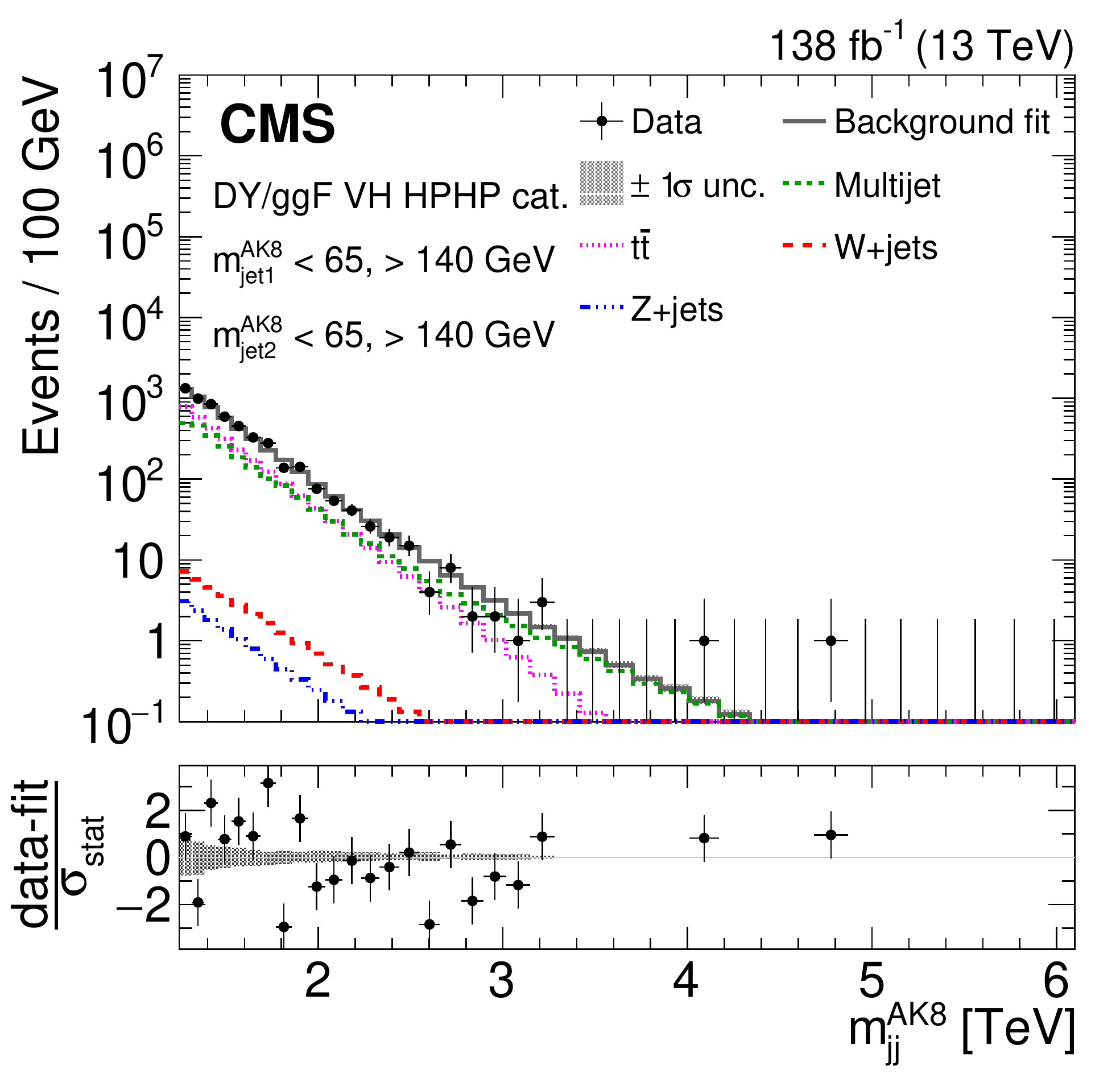}
\includegraphics[width=0.49\textwidth]{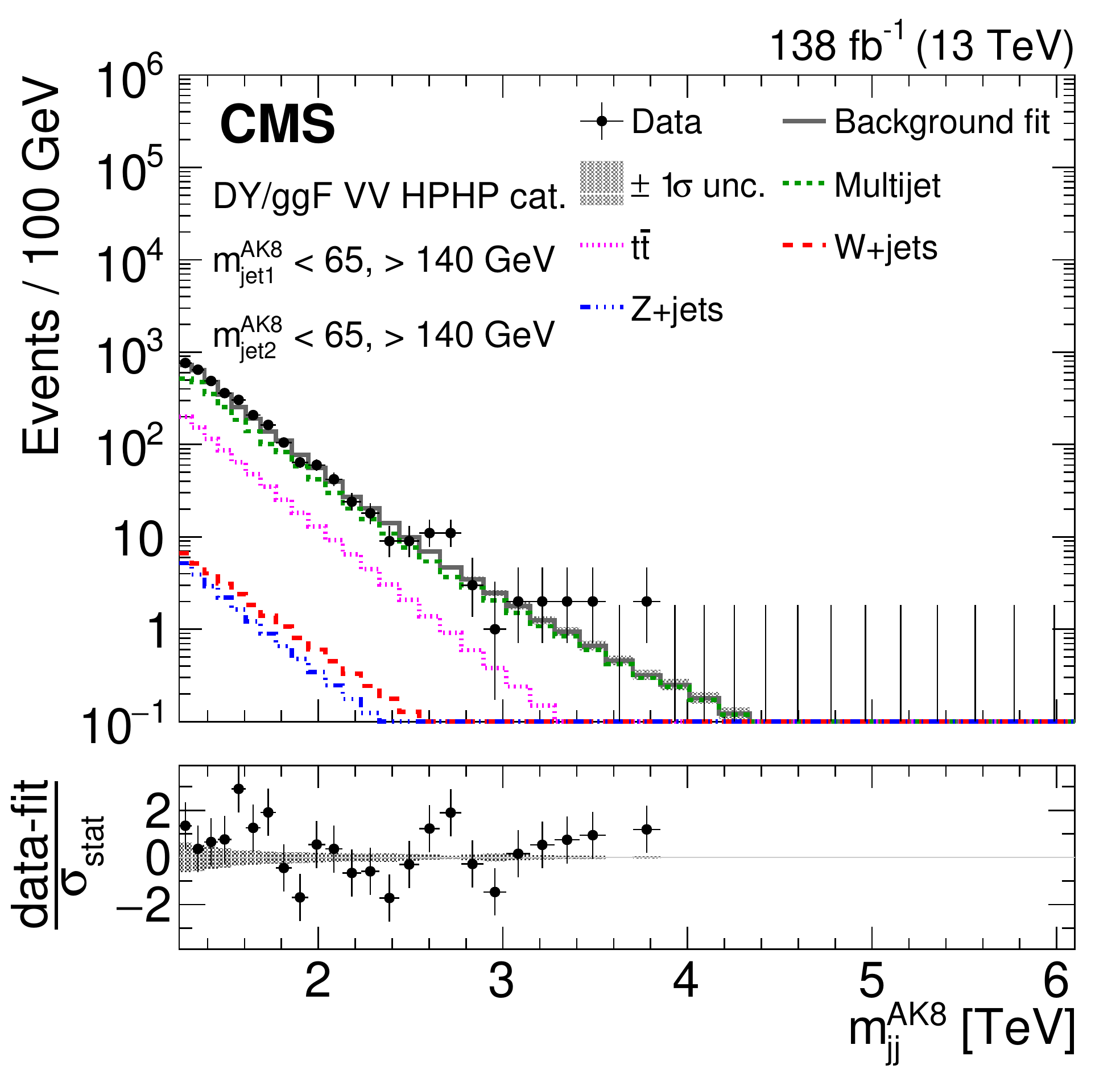}
\caption{Projections of the data and background post-fit distributions onto the \MVV dimension in regions enriched in background ($\MJO<65$ or $>$140\GeV, and $\MJT<65$ or $>$140\GeV), for the HPHP \VH (left) and \VV (right) DY/ggF categories. The background shape uncertainty is shown as a gray shaded band around the result of the maximum likelihood fit to the data under the background-only assumption (gray solid line), and the statistical uncertainties in the data are shown as vertical bars. The various background components contributing to the total background fit are also shown with different line colors.
Shown below each mass plot is the difference between the data and the fit divided by the statistical uncertainty in the data; the uncertainty bar represents the statistical uncertainty only. The total uncertainty in the background estimate fitted to the data divided by the statistical uncertainty in the data is shown as a band.}
\label{fig:postfit3}
\end{figure*}

Prior to the analysis of the full 3D spectra in data, the fit model was validated as follows.
The capability of the fit model to extract, without a significant bias, the cross section and significance of a potential signal resonance was confirmed with toy data distributions obtained from simulation with signal injected.
In data event samples enriched in QCD multijet and \ttbar events, obtained by excluding events with $65<\MJO<140\GeV$ and $65<\MJT<140\GeV$ in all categories of the analysis, a good description of the \MJO, \MJT, and \MVV was found.
Figure~\ref{fig:postfit3} shows an example validation for the \MVV spectra in data in the DY/ggF HPHP categories.
Finally, a goodness of fit test based on the saturated model is performed, comparing the full 3D spectra in data with the result of a maximum likelihood fit to the data under the background-only assumption. In this test, all uncertainties and their correlations are taken into account. A $p$-value of 0.35 is estimated w.r.t. the expected goodness of fit, indicating a good compatibility between the observed data and the background-only model.

\section{Results}
\label{sec:results}

\begin{figure*}[hbt!]
\centering
\includegraphics[width=0.49\textwidth]{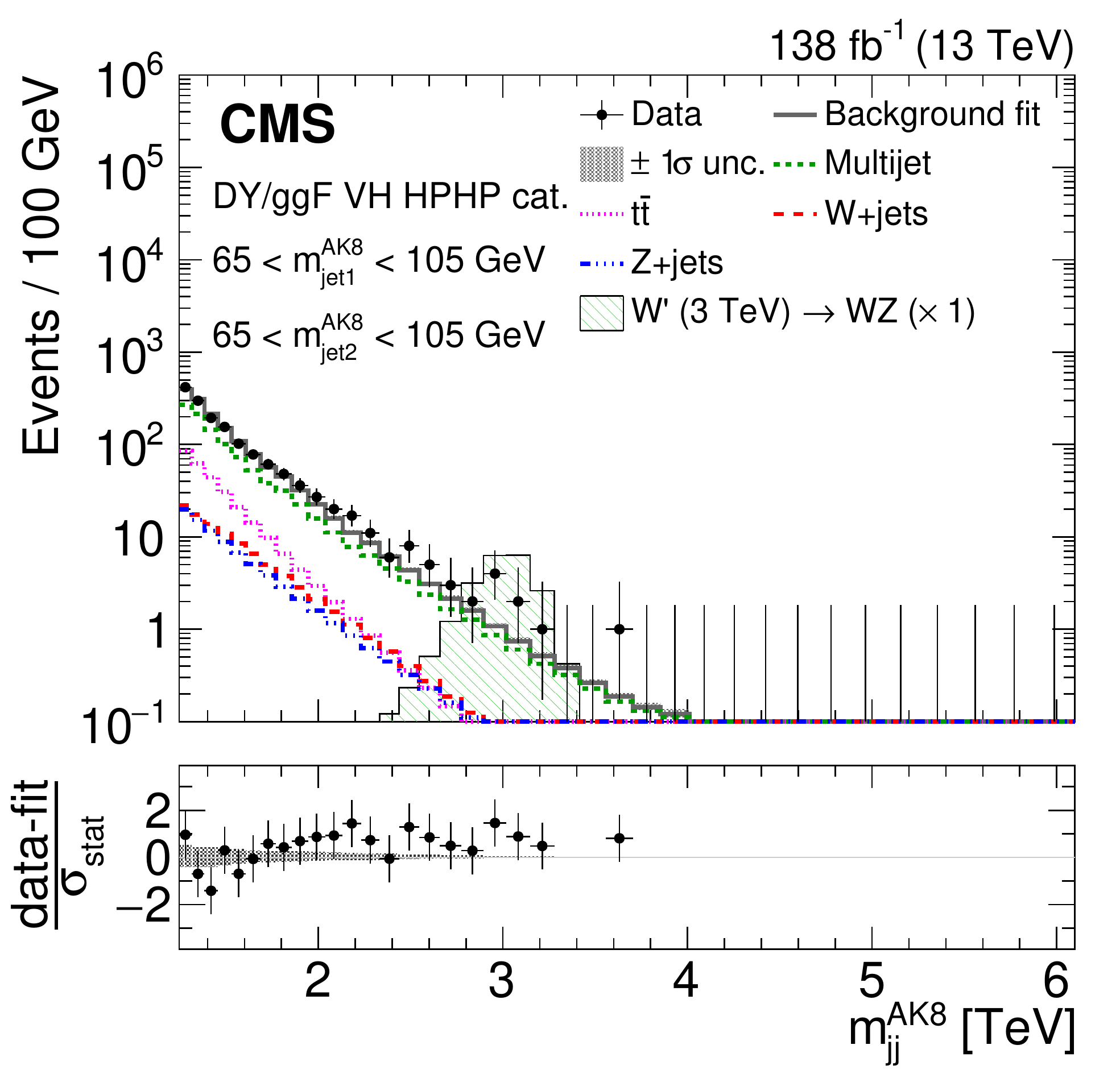}
\includegraphics[width=0.49\textwidth]{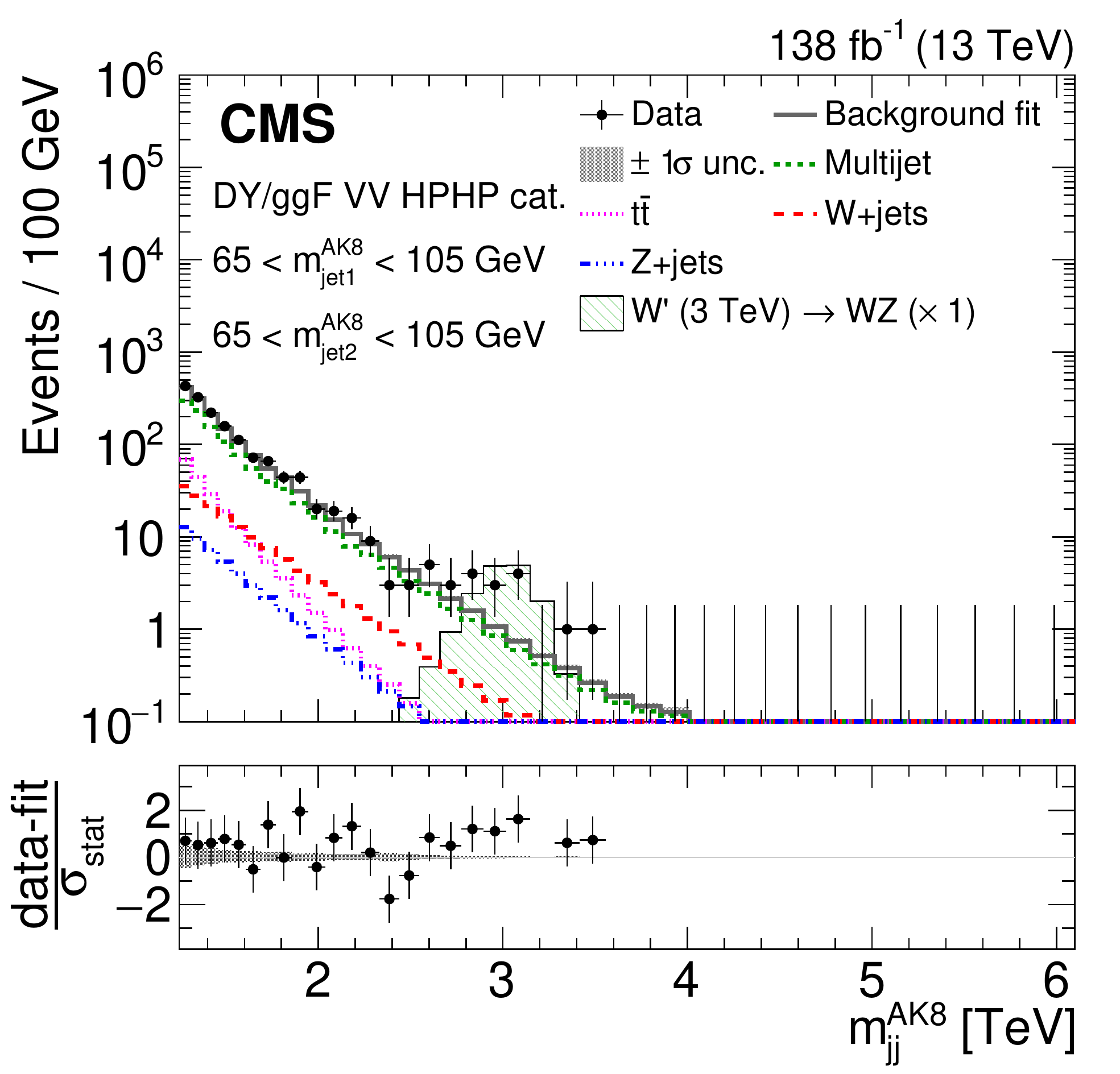}
\caption{Projections of data and background post-fit distributions onto the \MVV dimension in regions enriched in signal from DY/ggF \VV ($65<\MJO<105\GeV$, $65<\MJT<105\GeV$) for the HPHP \VH (left) and \VV (right) DY/ggF categories. The background shape uncertainty is shown as a gray shaded band around the result of the maximum likelihood fit to the data under the background-only assumption (gray solid line), and the statistical uncertainties in the data are shown as vertical bars. The various background components contributing to the total background fit are also shown with different line colors.
An example of a signal distribution is overlaid.
Shown below each mass plot is the difference between the data and the fit divided by the statistical uncertainty in the data; the uncertainty bar represents the statistical uncertainty only. The total uncertainty in the background estimate fitted to the data divided by the statistical uncertainty in the data is shown as a band.}
\label{fig:postfit4}
\end{figure*}

\begin{figure*}[htbp]
\centering
\includegraphics[width=0.49\textwidth]{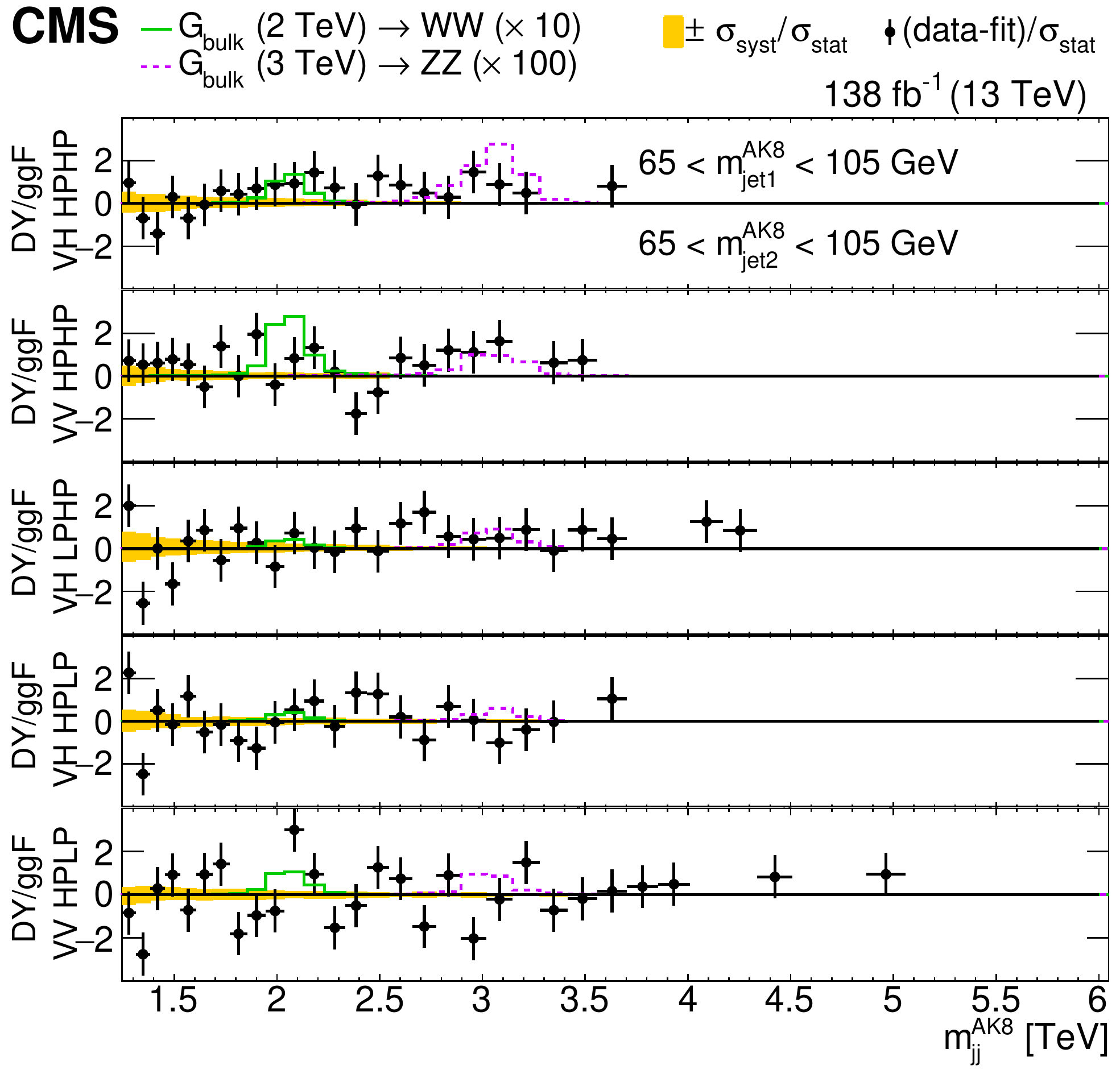}
\includegraphics[width=0.49\textwidth]{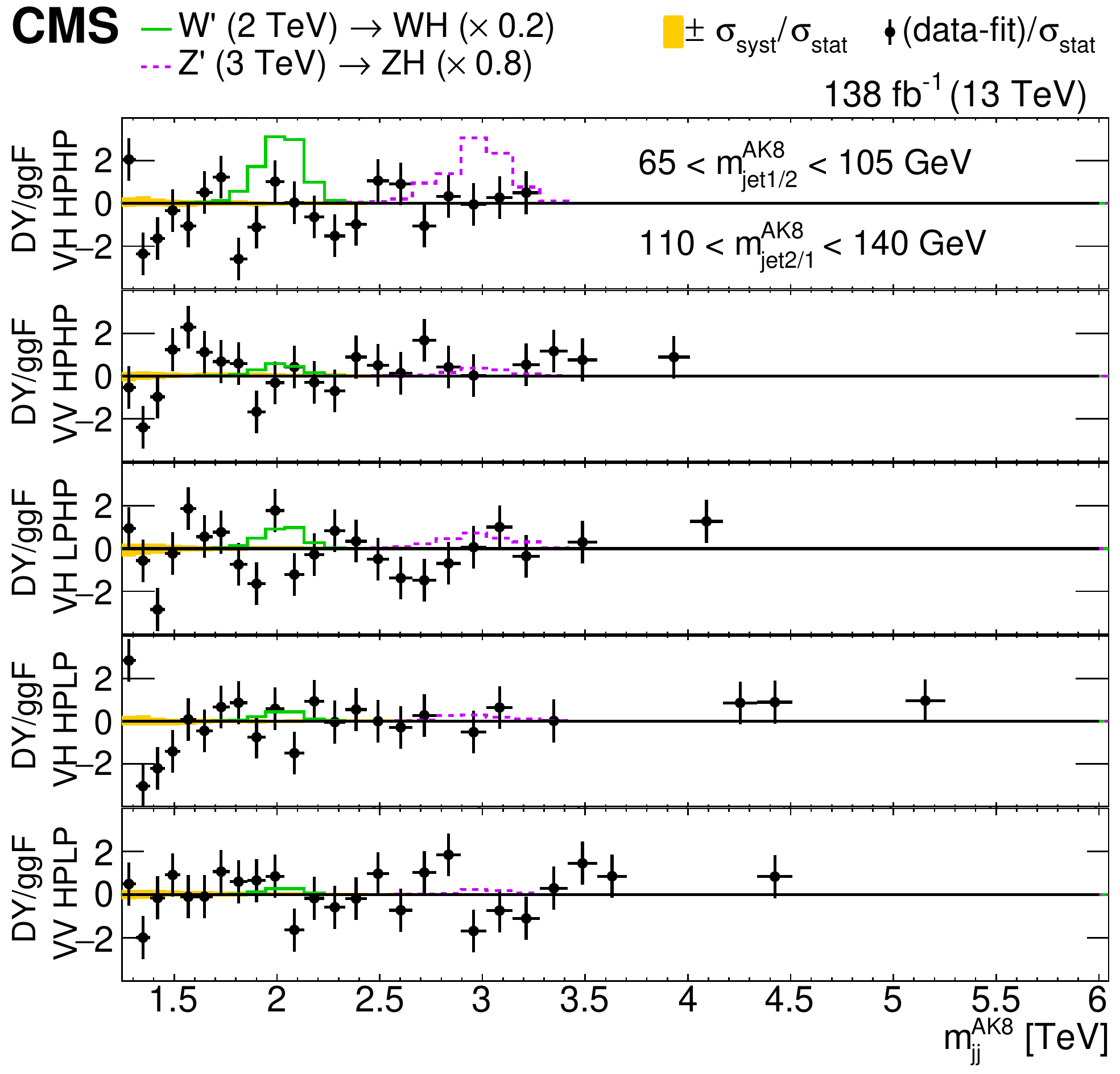} \\
\includegraphics[width=0.49\textwidth]{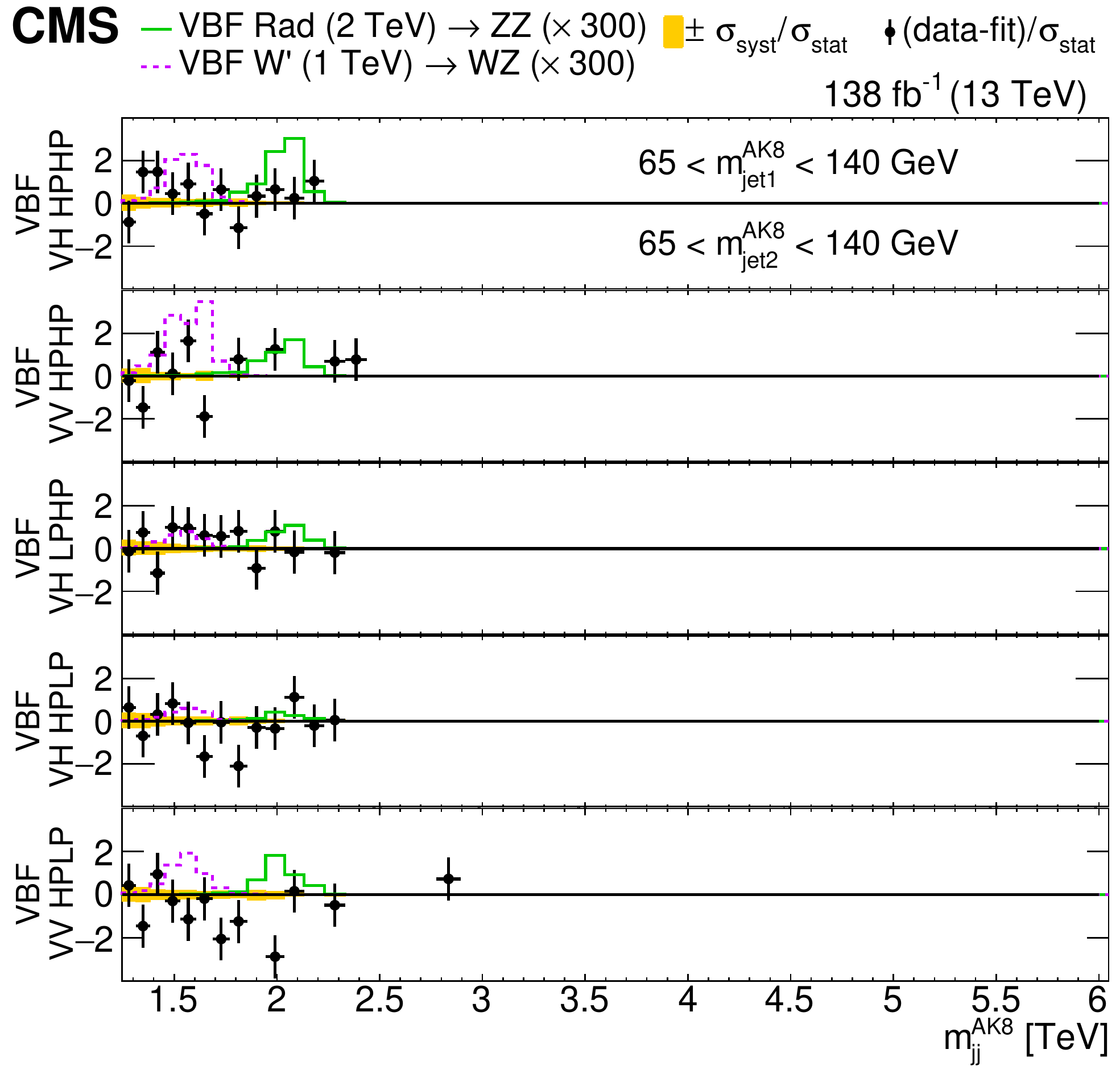}
\caption{Distributions of the difference between the data and the background-only fit divided by the statistical uncertainty in the data. The total uncertainty in the background estimate fitted to data divided by the statistical uncertainty in the data is shown as a band. It should be noted that these two uncertainties are partially correlated.
Three projections of the 3D phase space are shown in regions enriched in signal from DY/ggF \VV ($65<\MJO<105\GeV$, $65<\MJT<105\GeV$) (upper left), DY/ggF \VH ($65<\MJOT<105\GeV$, $110<\MJTO<140\GeV$) (upper right) and VBF \VV/\VH ($65<\MJO<140\GeV$, $65<\MJT<140\GeV$) (lower).
Examples of expected signal shapes added to the fit are overlaid, where the number of expected events is scaled by an arbitrary normalization factor stated in the legend.
}
\label{fig:postfit5}
\end{figure*}

Figure~\ref{fig:postfit4} shows the \MVV spectra in data for selected regions in \MJO and \MJT enriched in signal from DY/ggF \VV in the two categories most sensitive to the DY/ggF \VV signal.
The DY/ggF \VH HPHP category exhibits an excess of data events over the background prediction in the 1.7--3.2\TeV range, while in the DY/ggF \VV HPHP category excesses of data events around 2 and 3\TeV are visible.
In Fig.~\ref{fig:postfit5}, additional distributions of the difference between the data and the fit divided by the statistical uncertainty in the data in the \MVV spectra are shown for selected regions in \MJO and \MJT enriched in signal from DY/ggF \VV, DY/ggF \VH, and VBF \VV/\VH, representing the most sensitive regions of the 3D phase space.

For the statistical evaluation, we follow the \CLs prescription, using the asymptotic approximation described in Refs.~\cite{Junk:1999kv,CLs1,Cowan:2010js}. It was verified that limits computed with a full \CLs method agree with the asymptotic approximation approach within one standard deviation.

To check for the presence of a signal, we first compute asymptotic significances for all signal mass hypotheses in each signal scenario.
A maximum local significance of 3.6 standard deviations is observed at masses of 2.1 and 2.9\TeV under the $\PWpr\to\WZ$ hypothesis, corresponding to excesses in the spectra of the \VV HPLP and \VV HPHP event categories, respectively.
The size of the excess remains above two standard deviations when increasing significantly the \Vjets and \ttbar background uncertainties or using a looser (10\%) \DeepWZqq tagger requirement for the definition of the categories.
Taking into account the look-elsewhere effect, a global significance of 2.3 standard deviations is found.
Searches in the semileptonic final states~\cite{CMS:2021klu,Aad:2020ddw} did not observe any excesses at 2.1 and 2.9\TeV.

\begin{figure*}[htbp]
\centering
\includegraphics[width=0.4\textwidth]{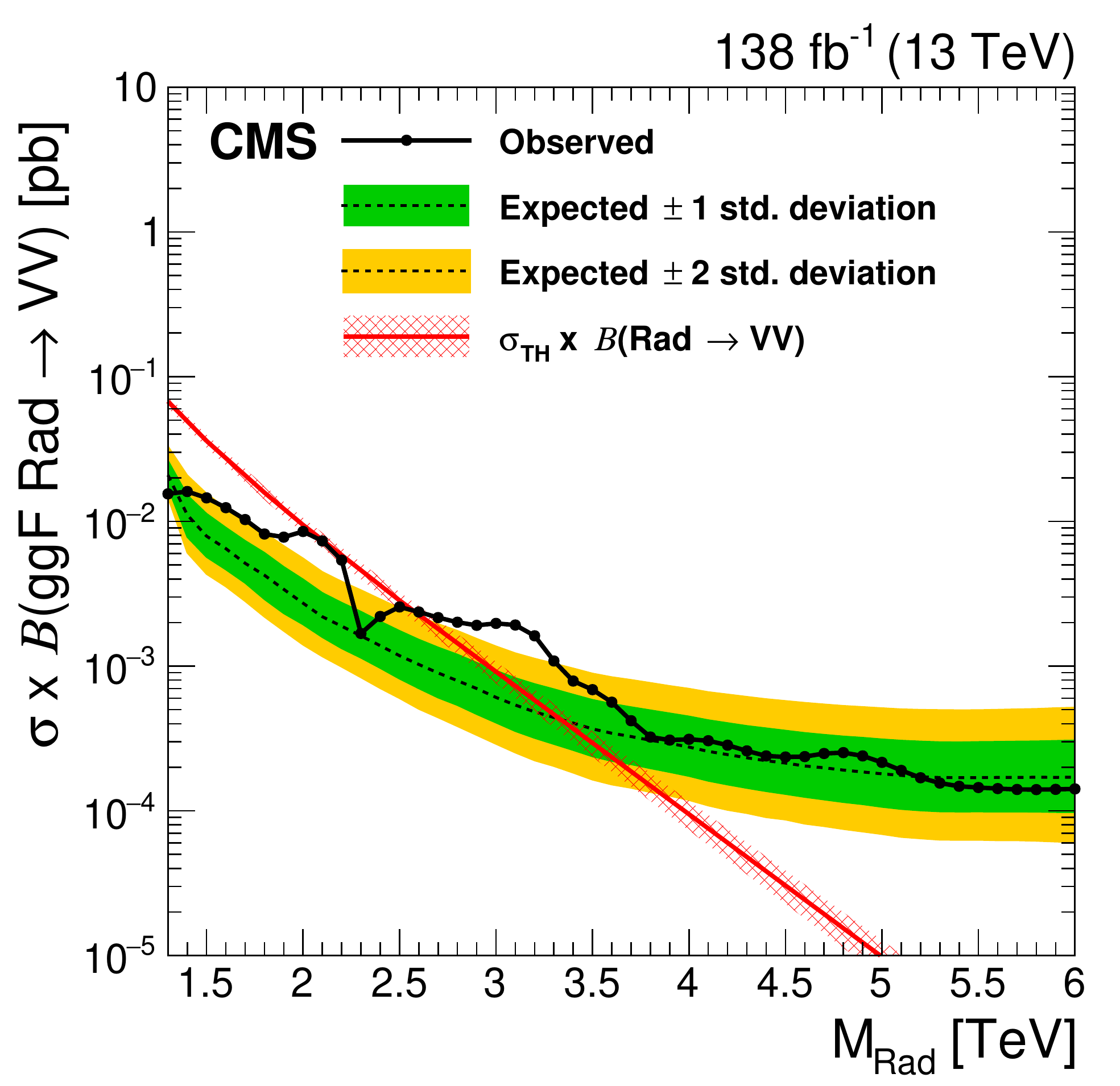}
\includegraphics[width=0.4\textwidth]{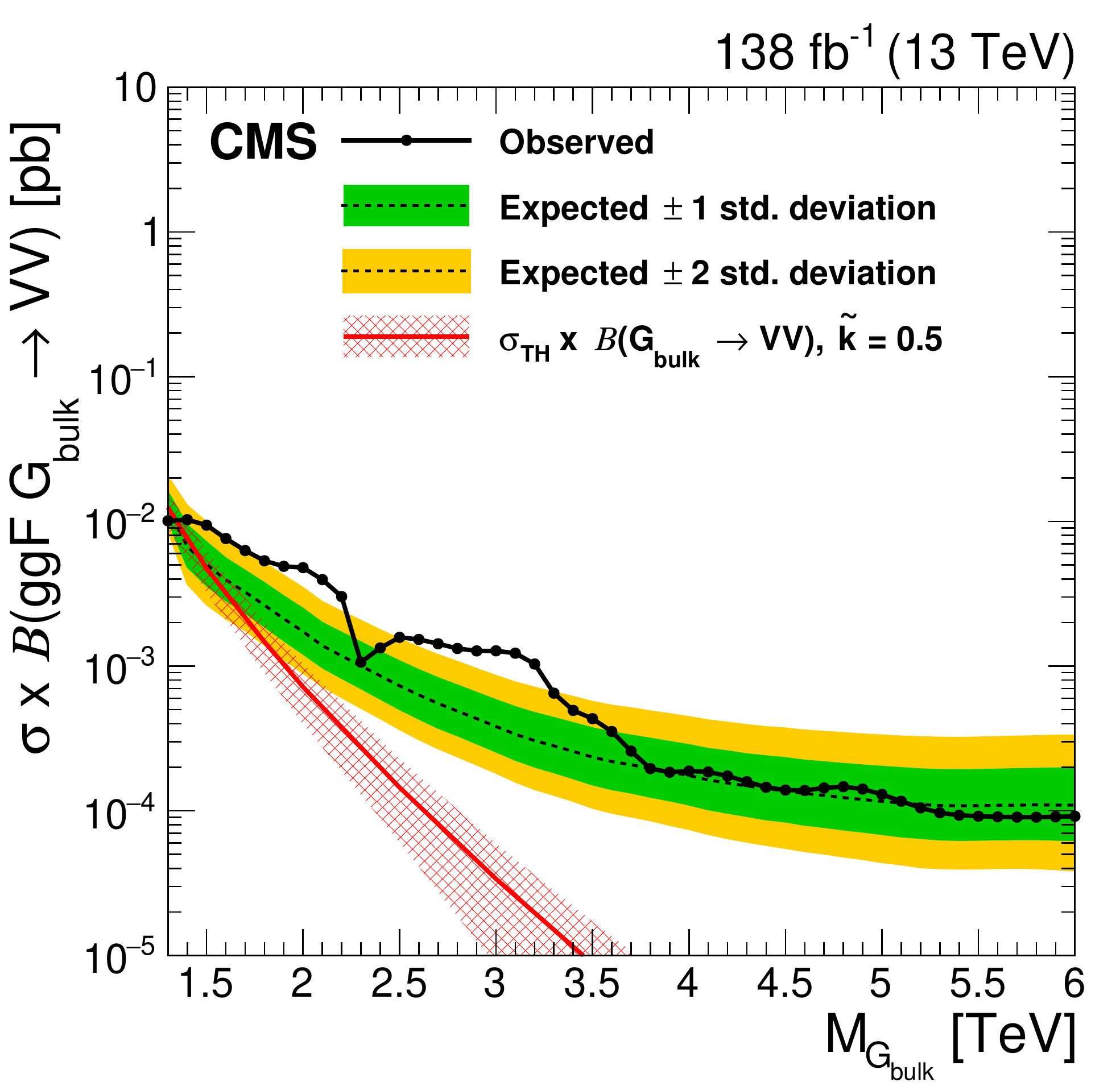}\\
\includegraphics[width=0.4\textwidth]{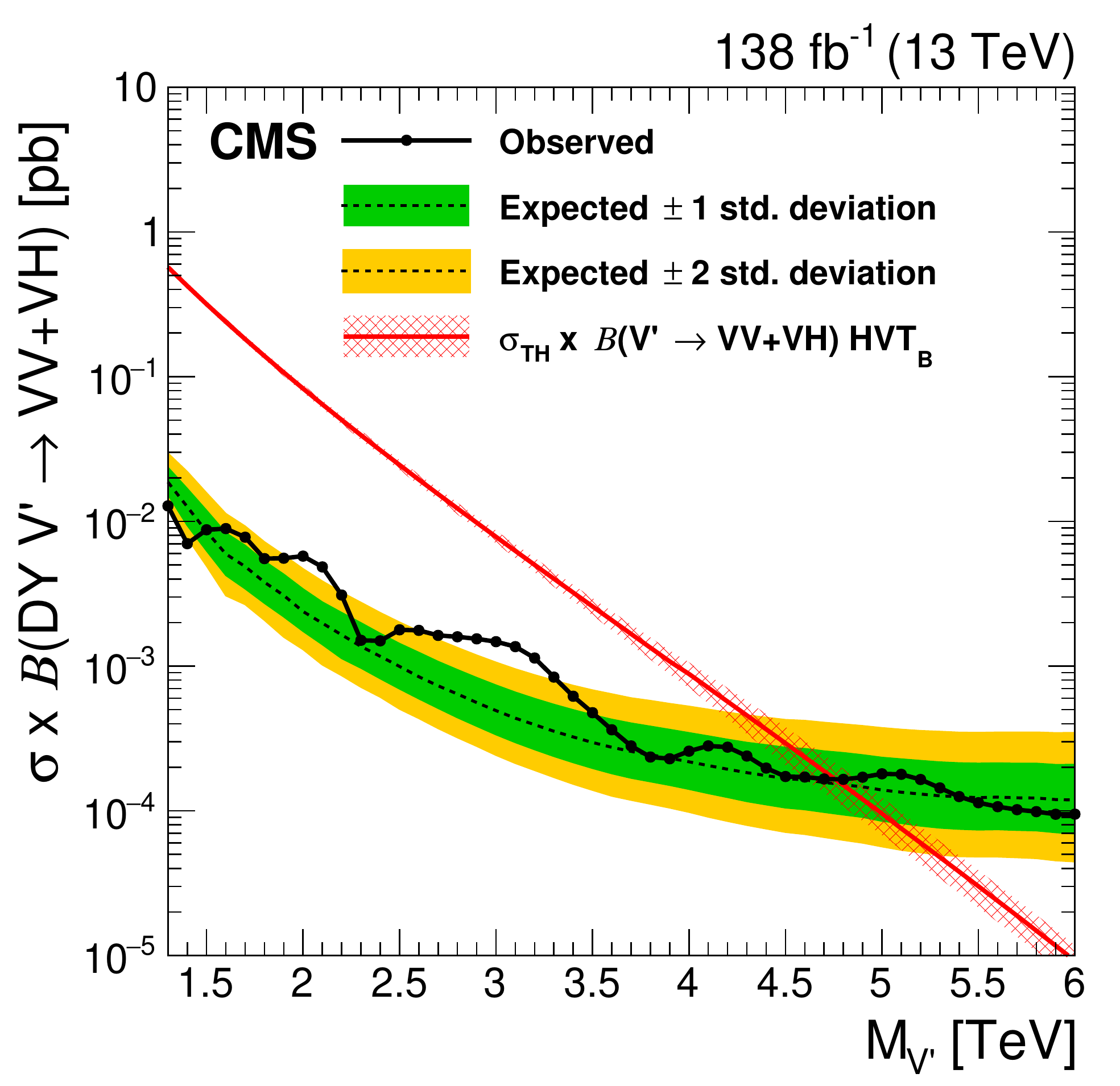}
\includegraphics[width=0.4\textwidth]{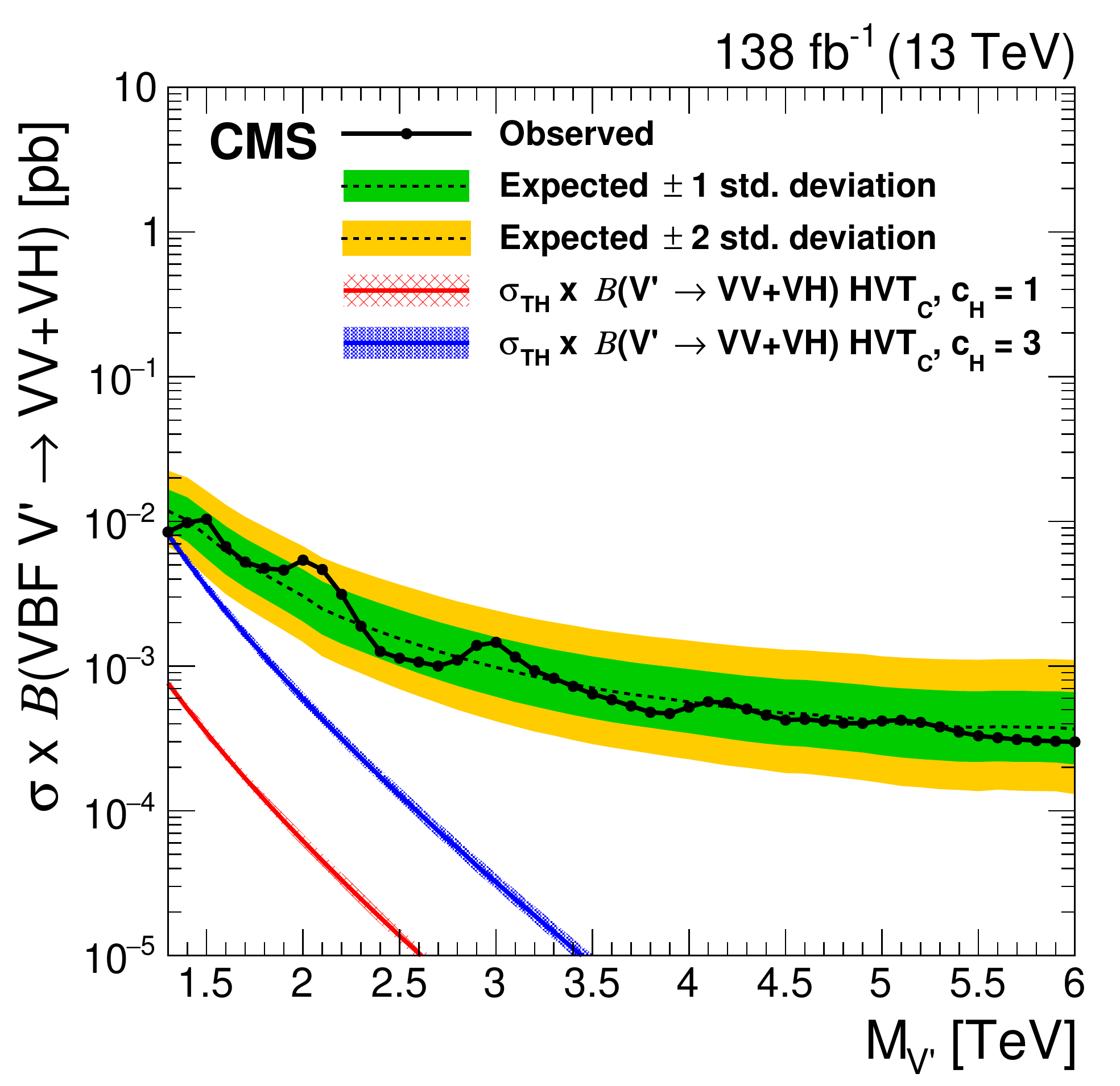}
\caption{Observed and expected 95\% \CL upper limits on the product of the production cross section ($\sigma$) and the branching fraction, obtained after combining all categories with \INTLUMI of data at $\sqrt{s}=13\TeV$, for $\Radion\to\VV$ (upper left), $\BulkG\to\VV$  (upper right), HVT model B $\PVpr\to\VV$+$\VH$ (lower left), HVT model C $\PVpr\to\VV$+$\VH$ (lower right) signals. For each signal scenario the theoretical prediction and its uncertainty associated with the choice of PDF set is shown.}
\label{fig:limits}
\end{figure*}

Upper exclusion limits on the production cross section at 95\% \CL are set.
Figure~\ref{fig:limits} shows the exclusion limits as functions of the resonance mass, compared to theoretical predictions of the product of the production cross section and the branching fraction $\mathcal{B}$.
The theoretical cross sections shown in Fig.~\ref{fig:limits} are calculated at NLO in QCD with the models detailed in Ref.~\cite{Oliveira:2014kla,Pappadopulo:2014qza}.
Table~\ref{tab:limits} summarizes the resonance mass limits at 95\% \CL for all considered models.
The limits presented in this letter are the most stringent to date in the all-jets final state.
The limits on spin-1 resonances decaying to a \WZ (\WH) boson pair in the HVT model B match or exceed by 0.1\TeV the previous published most stringent limits from the semileptonic final states~\cite{Aad:2020ddw,CMS:2021klu}.
Considering an HVT model with \VV and \VH resonances, this search sets the most stringent limit of \MassExclVPr.
The improvement by 15\% in the cross section exclusion limits for these resonances compared to the previous result by the CMS Collaboration in this channel~\cite{Sirunyan:2019jbg} arises in roughly equal measure from the increased size of the data set and the improved boson tagging techniques.
Tabulated results are provided in the HEPData record for this analysis~\cite{hepdata}.

\begin{table*}[htbp]
  \centering
  \topcaption{Summary of the exclusion limits on the resonance masses for the considered models. The numbers show the lower limit at 95\% \CL, with the exception of the ones given in parentheses quoting the ranges of exclusion.}
  \begin{tabular}{lccccccccc}
    Model & Decay channel & Observed limit (\TeV) & Expected limit (\TeV)\\
    \hline
    Radion ggF & \VV  &2.7  & 3.4 \\ [\cmsTabSkip]
    \BulkG ($\kappatilde=0.5$) ggF & \VV & 1.4 & 1.5\\ [\cmsTabSkip]
    HVT (B) \PWpr & \WZ & 4.4 & 4.3 \\
    HVT (B) \PWpr & \WH & 4.0 & 4.3\\
    HVT (B) \PZpr & \WW & (1.3--3.1, 3.3--3.5) & 3.8\\
    HVT (B) \PZpr & \ZH & 3.9 & 3.8\\
    HVT (B) \PVpr & \VV & 4.5 & 4.5 \\
    HVT (B) \PVpr & \VH & 4.2 & 4.5 \\
    HVT (B) \PVpr & \VV+\VH & 4.8 & 4.8 \\
  \end{tabular}
  \label{tab:limits}
\end{table*}

\ifthenelse{\boolean{cms@external}}{\clearpage}{}
\section{Summary}
\label{sec:summary}

A search has been presented for resonances with masses between \BulkGMassMinNoTeV and \BulkGMassMax that decay to \WW, \WZ, \ZZ, \WH, or \ZH boson pairs.
Each of the two boson decays is clustered into one large-radius jet, yielding a dijet final state from Drell--Yan and gluon fusion production, complemented by two additional jets for vector boson fusion production.
The hadronic decays of \PH, \PW, and \PZ bosons are identified using machine learning-based jet taggers that reduce the background from quantum chromodynamics multijet production.
No evidence of a departure from the expected background is found.
A maximum local significance of 3.6 standard deviations from the standard model prediction, corresponding to a global significance of 2.3 standard deviations, is observed at masses of 2.1 and 2.9\TeV.
Upper limits at 95\% confidence level on the resonance production cross section are set as a function of the resonance mass.
In a heavy vector triplet model, spin-1 \PZpr and \PWpr resonances with masses below \MassExclVPr are excluded.
These limits are the most stringent to date.
In a bulk graviton model, spin-2 gravitons and spin-0 radions with masses below \MassExclBulkG and \MassExclBulkR, respectively, are excluded.
Furthermore, the exclusive production of new heavy resonances through the vector boson fusion mode is constrained with upper cross section limits as low as \XsecExclVBF.

\begin{acknowledgments}
We congratulate our colleagues in the CERN accelerator departments for the excellent performance of the LHC and thank the technical and administrative staffs at CERN and at other CMS institutes for their contributions to the success of the CMS effort. In addition, we gratefully acknowledge the computing centers and personnel of the Worldwide LHC Computing Grid and other centers for delivering so effectively the computing infrastructure essential to our analyses. Finally, we acknowledge the enduring support for the construction and operation of the LHC, the CMS detector, and the supporting computing infrastructure provided by the following funding agencies: BMBWF and FWF (Austria); FNRS and FWO (Belgium); CNPq, CAPES, FAPERJ, FAPERGS, and FAPESP (Brazil); MES and BNSF (Bulgaria); CERN; CAS, MoST, and NSFC (China); MINCIENCIAS (Colombia); MSES and CSF (Croatia); RIF (Cyprus); SENESCYT (Ecuador); MoER, ERC PUT and ERDF (Estonia); Academy of Finland, MEC, and HIP (Finland); CEA and CNRS/IN2P3 (France); BMBF, DFG, and HGF (Germany); GSRI (Greece); NKFIH (Hungary); DAE and DST (India); IPM (Iran); SFI (Ireland); INFN (Italy); MSIP and NRF (Republic of Korea); MES (Latvia); LAS (Lithuania); MOE and UM (Malaysia); BUAP, CINVESTAV, CONACYT, LNS, SEP, and UASLP-FAI (Mexico); MOS (Montenegro); MBIE (New Zealand); PAEC (Pakistan); MES and NSC (Poland); FCT (Portugal); MESTD (Serbia); MCIN/AEI and PCTI (Spain); MOSTR (Sri Lanka); Swiss Funding Agencies (Switzerland); MST (Taipei); MHESI and NSTDA (Thailand); TUBITAK and TENMAK (Turkey); NASU (Ukraine); STFC (United Kingdom); DOE and NSF (USA).
    
\hyphenation{Rachada-pisek} Individuals have received support from the Marie-Curie program and the European Research Council and Horizon 2020 Grant, contract Nos.\ 675440, 724704, 752730, 758316, 765710, 824093, 884104, and COST Action CA16108 (European Union); the Leventis Foundation; the Alfred P.\ Sloan Foundation; the Alexander von Humboldt Foundation; the Belgian Federal Science Policy Office; the Fonds pour la Formation \`a la Recherche dans l'Industrie et dans l'Agriculture (FRIA-Belgium); the Agentschap voor Innovatie door Wetenschap en Technologie (IWT-Belgium); the F.R.S.-FNRS and FWO (Belgium) under the ``Excellence of Science -- EOS" -- be.h project n.\ 30820817; the Beijing Municipal Science \& Technology Commission, No. Z191100007219010; the Ministry of Education, Youth and Sports (MEYS) of the Czech Republic; the Hellenic Foundation for Research and Innovation (HFRI), Project Number 2288 (Greece); the Deutsche Forschungsgemeinschaft (DFG), under Germany's Excellence Strategy -- EXC 2121 ``Quantum Universe" -- 390833306, and under project number 400140256 - GRK2497; the Hungarian Academy of Sciences, the New National Excellence Program - \'UNKP, the NKFIH research grants K 124845, K 124850, K 128713, K 128786, K 129058, K 131991, K 133046, K 138136, K 143460, K 143477, 2020-2.2.1-ED-2021-00181, and TKP2021-NKTA-64 (Hungary); the Council of Science and Industrial Research, India; the Latvian Council of Science; the Ministry of Education and Science, project no. 2022/WK/14, and the National Science Center, contracts Opus 2021/41/B/ST2/01369 and 2021/43/B/ST2/01552 (Poland); the Funda\c{c}\~ao para a Ci\^encia e a Tecnologia, grant CEECIND/01334/2018 (Portugal); the National Priorities Research Program by Qatar National Research Fund; MCIN/AEI/10.13039/501100011033, ERDF ``a way of making Europe", and the Programa Estatal de Fomento de la Investigaci{\'o}n Cient{\'i}fica y T{\'e}cnica de Excelencia Mar\'{\i}a de Maeztu, grant MDM-2017-0765 and Programa Severo Ochoa del Principado de Asturias (Spain); the Chulalongkorn Academic into Its 2nd Century Project Advancement Project, and the National Science, Research and Innovation Fund via the Program Management Unit for Human Resources \& Institutional Development, Research and Innovation, grant B05F650021 (Thailand); the Kavli Foundation; the Nvidia Corporation; the SuperMicro Corporation; the Welch Foundation, contract C-1845; and the Weston Havens Foundation (USA).
\end{acknowledgments}

\bibliography{auto_generated}
\cleardoublepage \appendix\section{The CMS Collaboration \label{app:collab}}\begin{sloppypar}\hyphenpenalty=5000\widowpenalty=500\clubpenalty=5000\input{B2G-20-009-public-authorlist.tex}\end{sloppypar}
\end{document}

%% file: B2G-20-009-public-authorlist.tex
\cmsinstitute{Yerevan Physics Institute, Yerevan, Armenia}
{\tolerance=6000
A.~Tumasyan\cmsAuthorMark{1}\cmsorcid{0009-0000-0684-6742}
\par}
\cmsinstitute{Institut f\"{u}r Hochenergiephysik, Vienna, Austria}
{\tolerance=6000
W.~Adam\cmsorcid{0000-0001-9099-4341}, J.W.~Andrejkovic, T.~Bergauer\cmsorcid{0000-0002-5786-0293}, S.~Chatterjee\cmsorcid{0000-0003-2660-0349}, K.~Damanakis\cmsorcid{0000-0001-5389-2872}, M.~Dragicevic\cmsorcid{0000-0003-1967-6783}, A.~Escalante~Del~Valle\cmsorcid{0000-0002-9702-6359}, P.S.~Hussain\cmsorcid{0000-0002-4825-5278}, M.~Jeitler\cmsAuthorMark{2}\cmsorcid{0000-0002-5141-9560}, N.~Krammer\cmsorcid{0000-0002-0548-0985}, L.~Lechner\cmsorcid{0000-0002-3065-1141}, D.~Liko\cmsorcid{0000-0002-3380-473X}, I.~Mikulec\cmsorcid{0000-0003-0385-2746}, P.~Paulitsch, F.M.~Pitters, J.~Schieck\cmsAuthorMark{2}\cmsorcid{0000-0002-1058-8093}, R.~Sch\"{o}fbeck\cmsorcid{0000-0002-2332-8784}, D.~Schwarz\cmsorcid{0000-0002-3821-7331}, S.~Templ\cmsorcid{0000-0003-3137-5692}, W.~Waltenberger\cmsorcid{0000-0002-6215-7228}, C.-E.~Wulz\cmsAuthorMark{2}\cmsorcid{0000-0001-9226-5812}
% ...
\par}
\vskip\cmsinstskip
\dag:~Deceased\\
$^{1}$Also at Yerevan State University, Yerevan, Armenia\\
$^{2}$Also at TU Wien, Vienna, Austria\\
% ...
